\documentclass{aa} 

\usepackage{graphicx}
\usepackage{natbib}
\usepackage[colorlinks, citecolor=blue, linkcolor=blue]{hyperref}
\makeatletter
\renewcommand*\aa@pageof{, page \thepage{} of \pageref*{LastPage}} 
\makeatother

\usepackage{txfonts}
\newcommand{\gaia}{{\it Gaia}\xspace}
\newcommand{\chandra}{{\it Chandra}\xspace}

\begin{document}

\title{Coronal and chromospheric activity of Teegarden's star }

\author{B. Fuhrmeister\inst{\ref{inst1},\ref{inst2}}, J. H. M. M. Schmitt\inst{\ref{inst2}} 
\and A. Reiners\inst{\ref{inst4}} 
\and S. Czesla \inst{\ref{inst1},\ref{inst2}} 
\and V.~J.~S.~B\'ejar\inst{\ref{inst10},\ref{inst9}}
\and J. Caballero\inst{\ref{inst3}}, J. Eisl\"offel\inst{\ref{inst1}} 
\and Th. Henning\inst{\ref{inst5}}
\and J.~C.~Morales\inst{\ref{inst6},\ref{inst7}} 
\and A. Quirrenbach\inst{\ref{inst8}}
\and I. Ribas\inst{\ref{inst6},\ref{inst7}} 
\and J. Robrade\inst{\ref{inst2}}
\and P.~C. Schneider\inst{\ref{inst2}} 
\and M. Zechmeister\inst{\ref{inst4}}}

\institute{Th\"uringer Landessternwarte Tautenburg, Sternwarte 5, D-07778 Tautenburg, Germany\\
\email{bfuhrmeister@tls-tautenburg.de}\label{inst1} 
\and
        Hamburger Sternwarte, Universit\"at Hamburg, Gojenbergsweg 112, 21029 Hamburg, Germany\label{inst2}
\and
        Institut f\"ur Astrophysik, Friedrich-Hund-Platz 1,
        D-37077 G\"ottingen,
        Germany\label{inst4}
\and
        Instituto de Astrof\'{\i}sica de Canarias, c/ V\'{\i}a L\'actea s/n, E-38205 La Laguna, Tenerife, Spain\label{inst10}
\and
        Departamento de Astrof\'{\i}sica, Universidad de La Laguna, E-38206 Tenerife, Spain\label{inst9} 
\and
       Centro de Astrobiología (CSIC-INTA), ESAC campus, Camino bajo del castillo s/n, 28692 Villanueva de la Ca\~nada, Madrid, Spain\label{inst3}
\and    
        Max-Planck-Institute for Astronomy, K\"onigstuhl 17, D-69117 Heidelberg, Germany\label{inst5}
\and       
       Institut d'Estudis Espacials de Catalunya, E-08034 Barcelona, Spain\label{inst6}
 \and  
        Institut de Ci\`encies de l'Espai (CSIC), Campus UAB, c/ de Can Magrans s/n, E-08193 Bellaterra, Barcelona, Spain\label{inst7}
\and
        Landessternwarte, Zentrum f\"ur Astronomie der Universit\"at Heidelberg, K\"onigstuhl 12, D-69117 Heidelberg, Germany\label{inst8} 
        }

\date{Received dd/mm/2024; accepted dd/mm/2024}

\abstract{
    Teegarden's star is a late-type M-dwarf planet host, typically showing only rather low levels of activity. In this paper we present an extensive characterisation of this activity at photospheric, chromospheric, and coronal levels.  We specifically investigated TESS observations of Teegarden's star, which showed two very large flares with an estimated flare fluence between 10$^{29}$ and 10$^{32}$\,erg comparable to the largest solar flares.  We furthermore analysed nearly 300 CARMENES spectra and 11 ESPRESSO spectra covering all the usually used chromospheric lines in the optical from the \ion{Ca}{ii} H \& K lines at 3930\,\AA\, to the \ion{He}{i} infrared triplet at 10830\,\AA. These lines show different behaviour: The \ion{He}{i} infrared triplet is the only one absent in all spectra,  some lines show up only during flares, and  others are always present and highly variable. Specifically, the H$\alpha$ line is more or less filled in during quiescence; however, the higher Balmer lines are still observed in emission. Many chromospheric lines show a correlation with H$\alpha$ variability, which, in addition to stochastic behaviour, also shows systematic behaviour on different timescales including the rotation period.
    Moreover, we found several flares and also report hints of an erupting prominence, which may have led to a coronal mass ejection. Finally, we present X-ray observations of Teegarden's star (i.e. a discovery pointing obtained with the \emph{Chandra} observatory)
    and an extensive study with the \emph{XMM-Newton} observatory; when  these
    two large flares were observed, one of them showed clear signatures of the Neupert effect, suggesting the production of hard X-rays in the system.
    }
\titlerunning{Teegarden's star}
\authorrunning{Fuhrmeister et~al.}
\maketitle


\section{Introduction}

M dwarfs are the most common type of stars in our Galaxy;  as such, these stars are also expected to be the most common type of planet host star. This expectation is indeed verified by \citet{ribas2023}, who find a planet  occurrence rate of 1.44$\pm$0.20 planets per M dwarf using a sample of 
238 M dwarfs observed and monitored with the CARMENES instrument
\citep{Quirrenbach2020}.  Because of the very much reduced luminosity of M dwarfs, for example in comparison  to G-type stars, the habitable zones around M stars move far inwards, which allows the detection of planets in the
habitable zones around M dwarfs with conventional techniques.
For example, the late-type M dwarf TRAPPIST-1 (spectral type M 7.5) was found to host at least seven planets,  four of which are  in the habitable zone and all with Earth-like masses \citep{Gillon2017}.

However, M dwarfs exhibit ubiquitous magnetic activity, the signatures of which can be observed throughout the electromagnetic spectrum.  While 
magnetic activity is considered a nuisance by 
planet hunters, since it introduces noise in the radial velocity time series \citep{dumusque2011,Lafarga2023}, 
it is interesting in itself and with respect to the  solar-stellar connection \citep{Brun2017}. Specifically, along the  M-dwarf sequence the interior stellar structure changes since radiative cores  typical for Sun-like stars disappear  and the stars become fully convective.   Since the archetypal solar
$\alpha$ - $\Omega$ dynamos are located exactly at this interface between convection zone and radiative core, one may expect a change in the relevant dynamo process along the M-dwarf sequence \citep{giampapa1986}. Numerical simulations of fully convective stars demonstrate the production of dynamo action
\citep{Yadav2015}, and yet the observational evidence for activity differences between stars with and without radiative cores has remained somewhat elusive.  Finally, at the end of the M-dwarf sequence with very low photospheric temperatures, the mean ionisation level in the atmosphere decreases, leading in turn to a  reduced electric conductivity, which  may also affect the observed levels of activity \citep{Mohanty2003}.

In this paper we present a detailed activity study of Teegarden's star, an M7.0\,V star \citep{Alonso-Floriano2015} with low levels of activity, discovered only in 2003 by \citet{Teegarden2003} due to its faintness despite its proximity of 3.831$\pm$ 0.004\,pc \citep{Gaia2018A&A...616A...1G}. 
\cite{Marfil2021} redetermined the fundamental stellar parameters of Teegarden's star, which we list with other basic stellar and planetary parameters in Table~\ref{stellpar}.  
In addition, Teegarden's star was discovered to host at least three planets \citep{Zechmeister2019, Dreizler2024};  
the two inner planets,  Teegarden's star~b and c, are in (or at least close to) the habitable zone.

\begin{table}
    \caption{\label{stellpar} Parameters for Teegarden's star and its three planets.}
    \centering
    \begin{tabular}{@{}lcr@{}}
\hline
\hline
Parameter                   & Value        & Ref.\\
\hline
                            & Teegarden's star   & \\
Distance $d$ (pc)           & 3.831 $\pm$ 0.004  & \gaia DR2\\
Spectral type               & M7.0V              & Alo15 \\
$T_{\rm eff}$ (K)           & 3034 $\pm$ 45      & Marf21 \\
log\,g (cm/s)               &5.19 $\pm$ 0.2      & Marf21 \\
$\lbrack$Fe/H$\rbrack$                      & $-0.11$ $\pm$ 0.28   & Marf21 \\
$L_{\rm bol}$ ($ 10^{-5}L_\odot$)   & 72.2 $\pm$ 0.05 & Marf21 \\
$R$ ($R_\odot$)               & 0.120 $\pm$ 0.012 & Marf21 \\
$M$ ($M_\odot$)               & 0.097 $\pm$ 0.010 & Marf21 \\
log ($L_{{\rm H}\alpha}$/$L_{\rm bol}$)  & $-5.37$  & Zech19 \\
Age (Gyr)                   & $>$ 8  & Zech19 \\
$P_{\rm rot}$ (d)                  & 97.56   & Shan24\\
$L_{\mathrm{X}}$ (erg\,s$^{-1}$)               & $2.8...4.2\times 10^{25}$  & \\
$\log(L_{\rm X}$/L$_{\rm bol})$        & $-5.0$...$-4.81$   &\\
$F_{\rm X}$ (erg\,s$^{-1}$\,cm$^{-2}$)        & $1.58...2.4 \times 10^{-14}$ & \\
\hline
Planet b                    &        &  \\
Period (d)                  & 4.91   & Drei24\\
Semi major axis (au)        & 0.0259 & Drei24\\
\hline
Planet c                    &        &  \\
Period (d)                  & 11.416 & Drei24\\
Semi major axis (au)        & 0.0455 & Drei24\\
\hline
Planet d                    &        &  \\
Period (d)                  & 26.13 & Drei24\\
Semi major axis (au)        & 0.0791 & Drei24\\
\hline
    \end{tabular}
    \tablebib{
        Alo15:     \cite{Alonso-Floriano2015}
        Drei24:    \cite{Dreizler2024};
        \gaia DR2:     \cite{Gaia2018A&A...616A...1G};
        Marf21:     \cite{Marfil2021};
        Shan24:    \cite{Shan2024};
        Tee03:     \cite{Teegarden2003};
        Zech19:    \cite{Zechmeister2019}.
}
\end{table}


Our paper is structured as follows. In Sect.~\ref{sec:obs} we give an overview of the  data used, and we show the temporal placement of the individual observations in Sect.~\ref{sec:overview}. We analyse the TESS flares in Sect.~\ref{sec:TESS} and X-ray flares in Sect.~\ref{sec:xray}. The results of the optical data are presented in Sect.~\ref{sec:optical}. The impact of the  high-energy radiation on the planets is  discussed in Sect.~\ref{sec:habitability} and  the activity state of Teegarden's star is compared to other late-type M dwarfs in Sect.~\ref{sec:comparison}. In Sect.~\ref{conclusion} we present our conclusions.

\section{Observations and data analysis}\label{sec:obs}
The 298 CARMENES spectra used for the discovery of the planets also cover various chromospheric activity diagnostic lines, which we   study here in detail together with 11 ESPRESSO/VLT spectra.
Moreover, we present the results of two X-ray observations dedicated to Teegarden's star, which allowed us to assess the coronal activity properties of Teegarden's star, and finally we present three months of
space based optical photometry obtained with the TESS satellite. Together all these data allow us to study the variability of Teegarden's star at different wavelengths, and infer its influence on the two innermost planets.

\subsection{\emph{TESS}}

 While the prime scientific goal of the Transiting Exoplanet Survey Satellite (TESS) 
mission is the discovery of transiting exoplanets around brighter stars \citep{ricker2015}, the data is also well suited for stellar activity studies.  Short cadence TESS photometry (with a time resolution of two minutes) for Teegarden's star is available for
the TESS sectors 43 (2021-09-16 until 2021-10-11), 70 (2023-09-20  until  2023-10-16) and
71 (2023-10-16 until 2023-11-11).  All sectors were processed by the Science Processing Operations Center (SPOC) photometry and
transit-search pipeline \citep{Jenkins2016} and
we downloaded the respective light curves from the Mikulski Archive for Space Telescopes (MAST)\footnote{\tt{https://mast.stsci.edu}} for the simple aperture photometry (SAP).

\subsection{X-ray observations}

\subsubsection{\emph{Chandra Observatory}} 

We carried out a deep 50~ksec X-ray observation using the \chandra X-ray Observatory \citep{weisskopf1996} with its  High Resolution Imager Camera (HRC-I) in the focal plane \citep{murray1997}.  This instrumental setup is sensitive to X-rays between 0.1 - 10 keV, but it is important to realise that the HRC-I camera provides only extremely
limited energy resolution.
The observations were performed from  2019-10-03, UT 3:53 to UT 17:13 ($48092.2\,{\rm s}\approx13.3$\,hr) under the observation \chandra ObsId 22322 (PI: Schmitt).  Our data analysis was carried out with 
{\it Python} scripts working with the photon event lists as provided by the
standard pipeline.

\subsubsection{\emph{XMM-Newton}}

\emph{XMM-Newton} observed Teegarden's star on 2021 August 03 for 29.1\,ks.  The three X-ray CCD cameras pn, MOS1, and MOS2 (named after the kind of detector) of the European Photon Imaging Camera (EPIC) were used with
the thin filter inserted; due to the faintness of Teegarden's star
optical loading is irrelevant. They are all three operated in counting mode (i.e. they produce event lists stating characteristics of the detected photons such as arrival time, energy, and location).  The EPIC cameras are reasonably sensitive to photons in the energy range 0.4 keV - 7 keV, but the photons recorded from Teegarden's star are almost
all below 1 keV.
The optical monitor (OM) was used with
the U band filter in fast mode, thus providing time
resolution in the second range.  We downloaded the data as available in the
\textit{XMM-Newton} user archive under the sequence number 0883800101 and used
the \emph{XMM-Newton} Science Analysis System (SAS) 
\footnote{The \emph{XMM-Newton} SAS user guide can be found at  
\url{http://xmm.esac.esa.int/external/xmm\_user\_support/documentation/}} 
to reduce and analyse all \textit{XMM-Newton} data.

\subsection{Optical spectral data}

\subsubsection{CARMENES}
Most spectra used in this study were taken with the CARMENES spectrograph at the Calar Alto observatory, Spain \citep{Quirrenbach2020}.
CARMENES covers the wavelength range
from 5\,200 to 9\,600\,\AA\, (visual channel, VIS) and from 9\,600 to 17\,100\,\AA\,   (near-infrared channel, NIR) with a spectral resolution of
$\sim$ 94\,600 in VIS and $\sim$ 80\,400 in NIR.
CARMENES data are obtained mainly for planet search, nevertheless,
they are also a resource for studies of stellar
parameter determination and activity. Large parts of the CARMENES data (years $2016-2020$) have been made available publicly \citep{ribas2023}.

The stellar spectra were reduced using the CARMENES reduction pipeline
\citep{pipeline,Caballero2}. Subsequently, we corrected them for barycentric and
systemic radial velocity motions and carried out a correction for telluric absorption lines \citep{Nagel2023} using the {\tt molecfit}
package\footnote{\tt{https://www.eso.org/sci/software/pipelines/skytools /molecfit}}.

\subsubsection{ESPRESSO}
We also used 11 spectra taken with the Echelle SPectrograph
for Rocky Exoplanets and Stable Spectroscopic Observations (ESPRESSO) between 2019-09-21 and 2019-09-27, which we retrieved, reduced and flux-calibrated from the ESO archive.\footnote{\tt{https://archive.eso.org/}} The resolution of the spectra is 140\,000, and they cover the range \mbox{3800 – 7900\,\AA;} this means that  ESPRESSO does cover the \ion{Ca}{ii} H \& K lines, which are not covered by CARMENES, but are important for activity studies of solar-like stars \citep{Baliunas1995, SM2017, Perdelwitz2024}.

\subsubsection{Pseudo equivalent width measurements}
To assess the activity state of Teegarden's star in each spectrum, we employed pseudo-equivalent width (pEW) measurements, since late M dwarfs do not show an identifiable continuum because of the abundance of molecular absorption lines. For the used central wavelength, full width of the line integration window, the location of the two reference bands and a detailed description of pEW measurements of chromospheric lines, we refer to \citet{cycle, Fuhrmeister2023}.


\section{Overview}\label{sec:overview}
\begin{figure*}
  \hspace{-1.5cm}
  \includegraphics[width=1.15\textwidth, clip]{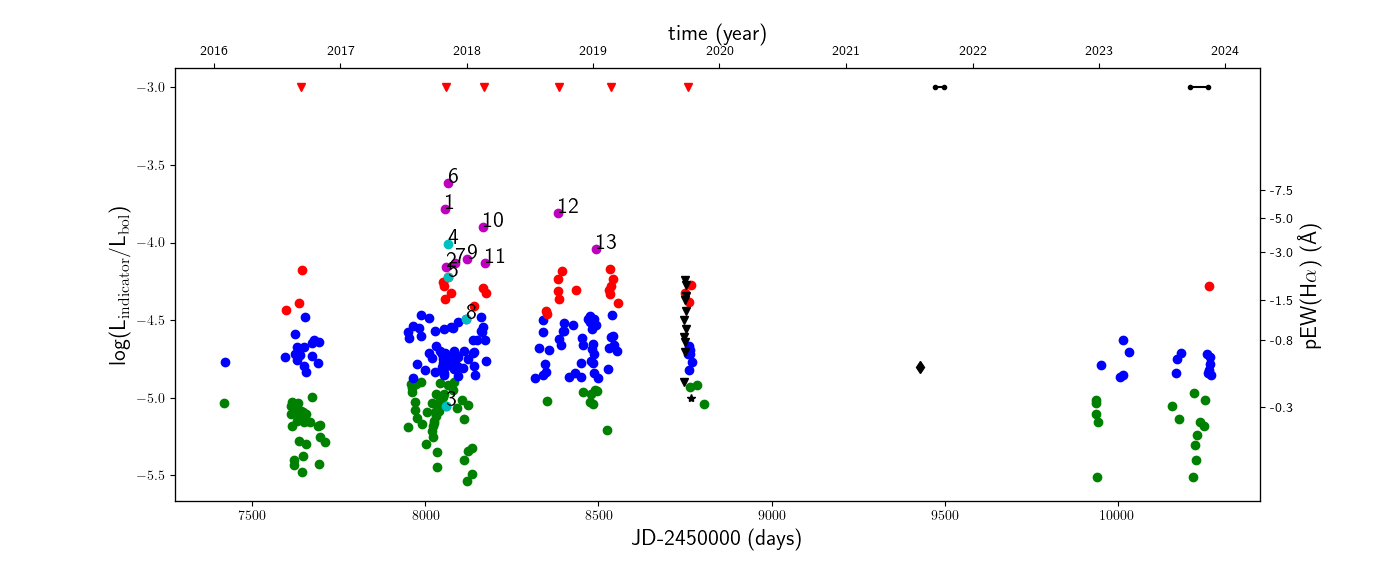}
\begin{center}
\caption{\label{pEWhalpha} Time series of $L_{\mathrm{indicator}}/L_{\mathrm{bol}}$. For the CARMENES 
  $L_{\mathrm{H}\alpha}/L_{\mathrm{bol}}$ we mark low activity states as green and blue dots, high activity states as  red dots, flares as magenta dots, and spectra with an asymmetric H$\alpha$ shape as cyan dots (see Sect.~\ref{sec:halpha} for a detailed discussion). The cyan and magenta dots are labelled with the flare number also used in Fig.\ref{subtracspec}. Flare no. 14 is not shown here since the spectrum gets into absorption, and therefore no $L_{\mathrm{indicator}}/L_{\mathrm{bol}}$ can be calculated with the $\chi$ method used here since it is only defined for emission lines.
  The ESPRESSO H$\alpha$ measurements are marked as black triangles.
  The $L_{\rm X}/L_{\mathrm{bol}}$ measurement of the \emph{Chandra} observation is marked as a black star; that of the \emph{XMM-Newton} observation is marked as a black diamond.  The time spans of the TESS observations are marked as   small black dots connected by a black line. Since TESS is not photometrically calibrated, the position on the y-axis is arbitrary (we note here that for a blackbody of the temperature of Teegarden's star about 20 percent of the radiation is in the TESS band). The red triangles mark the positions of the clusters of the higher activity states.
}
\end{center}
\end{figure*}

The observations of Teegarden's star were not part of a dedicated multi-wavelength campaign; nevertheless some of the data were taken (quasi-) simultaneously. To provide a better overview of the data, we show all used observations in Fig.~\ref{pEWhalpha}. H$\alpha$ pEWs are converted to $L_{\mathrm{H}\alpha}/L_{\mathrm{bol}}$-values by using the $\chi$-factor following the calculation by \citet{Reiners2008}. Our $L_{\mathrm{H}\alpha}/L_{\mathrm{bol}}$-values agree with calculations by \citet{Zechmeister2019}. The $L_{\rm X}/L_{\mathrm{bol}}$-values are calculated in Sect.~\ref{sec:xray}. 

Regarding the simultaneity of the data, we note that the \emph{XMM-Newton} observation took place during the long observation gap of CARMENES, when also the TESS observations of sector 43 were obtained.
The TESS sector as well as the \emph{XMM-Newton} observation contains two significant flares (see Sect.~\ref{sec:TESS} and \ref{sec_xmm_flare} for a detailed discussion).
The TESS observations of sectors 70 and 71 cover 8 CARMENES spectra. These TESS observations show no apparent flare activity. Also the CARMENES pEW(H$\alpha$) values suggest a relatively low activity state for all of these spectra (for an empirical definition of the low activity state of Teegarden's star regarding the H$\alpha$ line see Fig.~\ref{pEWhalpha} and Sect.~\ref{sec:halpha}). The TESS observation of sector 71 ended 5 days before the next CARMENES spectrum, which was then in high activity state. The two adjacent TESS light curves could in principle be searched for signatures of the rotation period, which is with 97.6 days longer than even two TESS sectors, but no indications of rotational modulation can be identified.

The short timescales of the variations are demonstrated by the \emph{Chandra} observation, which occurred only 5.5 hours after a CARMENES observation in high activity state. Nevertheless, it appears to cover the quiescent state of Teegarden's star since the X-ray flux from the quiescent phases of the \emph{XMM-Newton} observation is very similar (see Sect.~\ref{sec:xray}). We therefore argue that this CARMENES observation may have covered a small flare or the decay phase of a larger flare, which ended before the \emph{Chandra} observation started.

\section{Flaring activity observed by TESS}\label{sec:flaring}

\subsection{TESS flare light curves}\label{sec:TESS}

  We examined the light curves obtained in all three sectors, one from 2021, two from 2023.  In the two sectors recorded in
2023 we could not find any obvious flares, while  in 2021 two 
very obvious flares were recorded by TESS, which we call Flare I and Flare II for reference. We note that we are not saying that there 
are no more flares, but rather that any other flares are weaker and
increasingly difficult to distinguish from the photometric noise.

\begin{figure*}[t]

    \includegraphics[width=0.5\textwidth]{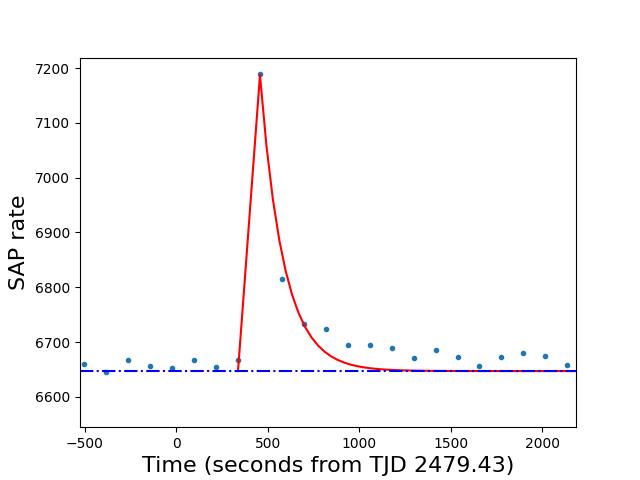}
    \includegraphics[width=0.5\textwidth]{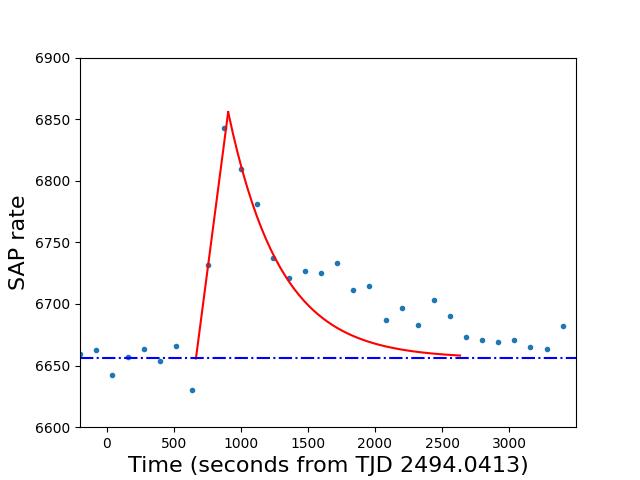}
    \caption{\label{fig:tess_fl_1}
TESS light curve of Flare I (\emph{left}) and Flare II (\emph{right}). The TESS data points are shown in blue, a simple analytic model is shown in red, and the assumed background is the blue dash-dotted line.
}
\end{figure*}

The TESS short cadence data have a time resolution of two minutes, which is too
long to adequately record flare light curves. An inspection of the TESS light curves
(Fig.~\ref{fig:tess_fl_1}) shows that 
very few data points are actually ascribable to the flares.
We therefore abstain from a detailed 
light curve modelling, rather we describe the TESS light curves with a simple linear
rise followed by an exponential decay; already this model requires five 
parameters: the constant background level, the times of start and peak of the flare, the peak amplitude, and the decay time.

In Fig.~\ref{fig:tess_fl_1} (left panel), we show the light curve of Flare I and a simple model curve.  In Fig.~\ref{fig:tess_fl_1} it can be seen that
the flare rise is actually unresolved.  Since the flare rise is usually much faster than the flare decay, the flare rise was probably shorter than 2 minutes and the flare amplitude larger than the largest recorded flux.  A simple exponential decay does not adequately describe the full flare light curve, there appears to be some ``fading out'' on longer timescales.

Flare II looks similar to Flare I, and yet there are some differences: At least one data point is recorded during the flare rise, the overall amplitude is smaller than for Flare I, and yet the flare decay time is considerably longer. Again a simple
exponential decay  is  an inadequate description of the
flare light curve.





\subsection{Energetics of TESS flares}
\label{sec_flare_tess_ener}

In the following section we provide  estimates of the energetics of Flare I and II shown in Fig.~\ref{fig:tess_fl_1}.
These estimates are only rough estimates, since the temperature of the flaring material and its temporal evolution are unknown and cannot be deduced from the TESS data.  For this reason it is
also not worthwhile using actual flare model atmospheres, rather, following the approach applied by \cite{shibayama2013},
we take recourse to simple blackbody models to describe the energetics of
both the photospheric and flaring emission.   One must bear in mind
that the physical processes during a stellar flare are rather 
complicated and that stellar flares can have vastly different individual properties; for a detailed recent review of stellar flares and in particular flare modelling we refer to \cite{kowalski2024} and references therein.

To calibrate the recorded TESS light curves we use the stellar parameters listed in
Table~\ref{stellpar}.  Using the stellar distance and the effective 
temperature, and the
TESS transmission curve we can compute the incident energy flux and thus derive the conversion factor to convert the observed TESS SAP rate into energy flux.   Using this conversion
factor also for the flare SAP rate and with some 
assumed effective temperature of the flaring material, we can compute the flare energy flux.   Using the peak flare 
amplitude one can then determine the emitting area, and with the decay time, the
total optical flare energy.

\begin{table}
    \caption{\label{flarepartess} Flare parameters for Teegarden's star from TESS light curves}
    \centering
    \begin{tabular}{@{}lrr@{}}
\hline
\hline
Parameter                   & Flare I        & Flare II\\
\hline
Time of flare peak (TJD)    & 2479.440       & 2494.051\\
SAP counts in flare decay    & 70000          & 78000 \\
SAP counts in flare rise     & 32000          & 24000 \\
Total SAP flare counts       & 102000         & 102000 \\
SAP peak count rate (counts/s)        & 540            & 200 \\
Decay time (s)            & 130            & 390 \\
\hline
$T_{\rm flare}$ = 15000~K       &                &\\
Peak flare flux (erg\,s$^{-1}$)     & $7.1\times 10^{29}$  &$2.6\times 10^{29}$ \\
Total flare energy (erg)    & $1.3\times10^{32}$  &$1.3\times 10^{32}$  \\
Flare area (cm$^2$)         & $2.5\times10^{17}$  & $9.2\times 10^{16}$ \\
\hline
$T_{\rm flare}$ = 8000~K       &                &\\
Peak flare flux (erg\,s$^{-1}$)     & $2.3\times 10^{29}$  &$8.0\times 10^{28}$ \\
Total flare energy (erg)    & $4.3\times 10^{31}$  &$4.3\times 10^{31}$  \\
Flare area (cm$^2$)         & $9.9\times 10^{17}$  & $3.7 \times10^{17}$ \\
\hline
    \end{tabular}
\end{table}

In Table~\ref{flarepartess} we list the parameters derived in this fashion.   The physical flare parameters were computed by assuming
temperatures of 15000~K and 8000~K for the flaring plasma; since the
flaring plasma is expected to change its temperature during the
flare evolution, we may expect that the quantities listed in
 Table~\ref{flarepartess} provide reasonable estimates, which allow to
 put Flare I and II into a physical context.   While there is
 admittedly quite some uncertainty w.r.t. the derived total flare
 energies, the numbers presented in Table~\ref{flarepartess} suggest
 that the flare energies involved are comparable to the
 flare energies of the largest solar flares observed (see
 \citet{moore2014} for a discussion of total solar irradiance measurements of
 solar flares).

\citet{Seli2021} computed flare frequency distributions for very late-type M dwarfs observed with TESS. Inserting our flare energies in their Equ. 12 leads to an expectation of a flare like the ones we observed every 180 or 80 days for our largest and lowest flare energy, respectively. This is in rough agreement with our overall flare frequency of 2 flares in about 80 days which leads to 2.6$\pm$1.8 flares in 100 days. The occurrence of the two flares in just one sector, however, may  hint at more active times of Teegarden's star (cf. Sect.~\ref{sec:timing}).

\section{X-ray and UV observations with {\it Chandra} and {\it XMM-Newton}}\label{sec:xray}

\subsection{{\it Chandra} X-ray source detection and flux determination}

An analysis of the counting events received from the
central source agrees with the position of Teegarden's star to better than one arcsec, by taking the proper motion of the star into account.
To determine the total number of counts recorded from Teegarden's star, we determine the cumulative
number of counts (CNC) in concentric rings centred on the observed source position.   For the case of uniform background, CNC should increase quadratically once no source photons contribute any longer. For the offset of the quadratic fit we find a value of 92.9 $\pm$ 1.7, which represents our best estimate for the recorded number of source counts from Teegarden's star with a Poisson uncertainty of 9.8,
which then translates into a (dead-time corrected) count rate of (1.96 $\pm$ 0.21)\,10$^{-3}$\,cts/s.

To convert the observed count rate into an energy flux, we need to multiply with the energy-to-count
conversion factor (ECF), which, however, depends on the assumed spectral parameters.  The HRC data
provide only very little energy resolution, hence no formal spectral fits are possible.   
Using PIMMS at NASA's HEASARC \cite{mukai1993}, we
can compute ECFs to convert the observed count rates 
To remedy this situation, we use
temperature estimates from \textit{XMM-Newton}.  For ''cool" temperatures in the range log~T between 6.2 and 6.5 and
abundances between solar and 0.4 sub-solar, these ECFs change very little (less than 1\%), while larger
changes occur for lower or higher temperatures.   Using then an ECF of
$1.09 \times 10^{-11}$\,erg\,ct$^{-1}$\,cm$^{-2}$,  leads to an
X-ray flux of $2.14 \times 10^{-14}$ erg\,s$^{-1}$\,cm$^{-2}$.   With this flux
one then computes -- using the information in Table~\ref{stellpar} --
an X-ray luminosity $L_{\rm X}$ of $3.7 \times 10^{25}$ erg\,s$^{-1}$ and a logarithmic $L_{\rm X}/L_{bol}$-ratio of -4.9.

\subsection{{\it Chandra} X-ray variability}

\begin{center}
    \begin{figure}
        \includegraphics[width=9.0cm,height=6cm]{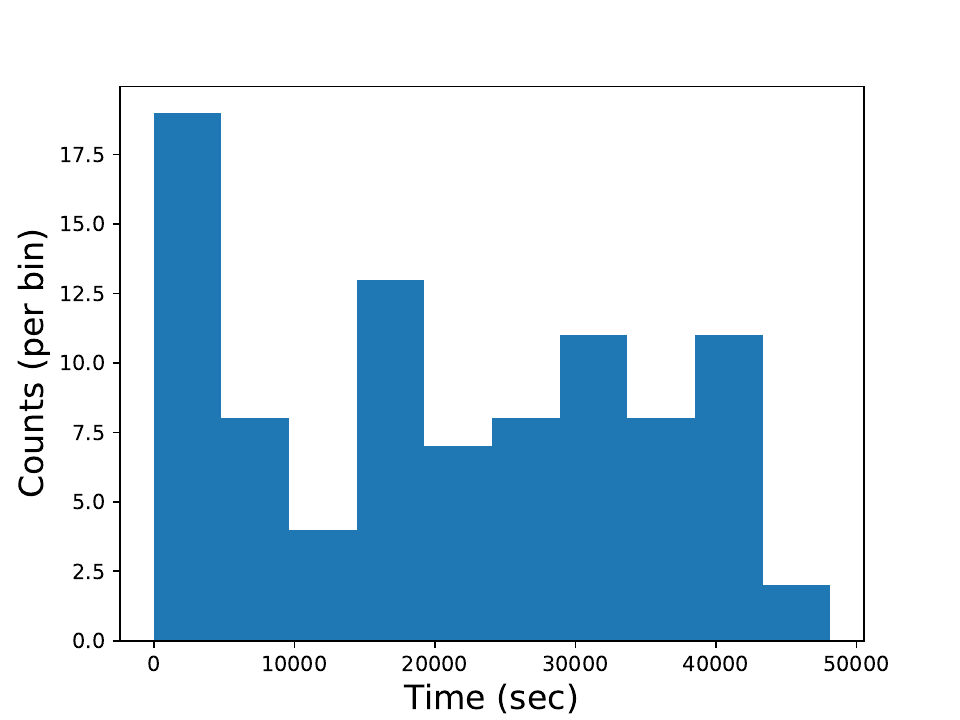}
        \caption{\label{fig4}
        Binned \chandra light curve of Teegarden's star with time bins of $\approx$ 2500\,s.}
    \end{figure}
\end{center}

We next considered the temporal behaviour of the X-ray emission from Teegarden's star as recorded by {\it Chandra}. To this end we constructed binned light curves using different
choices of temporal binning, where the two highest bins occur adjacent to another right at the beginning of
the observations (see Fig. \ref{fig4}). Performing a $\chi^2$ test we find variability at the significance level of $\approx$ 98\%,
somewhat depending on the precise choice of bins. 
A Kolmogorov-type analysis of the photon arrival times
shows variability only at a slightly lower confidence level (90\%).
Thus we conclude that the X-ray emission of Teegarden's star during the {\it Chandra} observations shows no larger flares, although some
low-level variability is possibly present in the form of weaker
flares.

\begin{figure}[h]

    \includegraphics[width=0.5\textwidth]{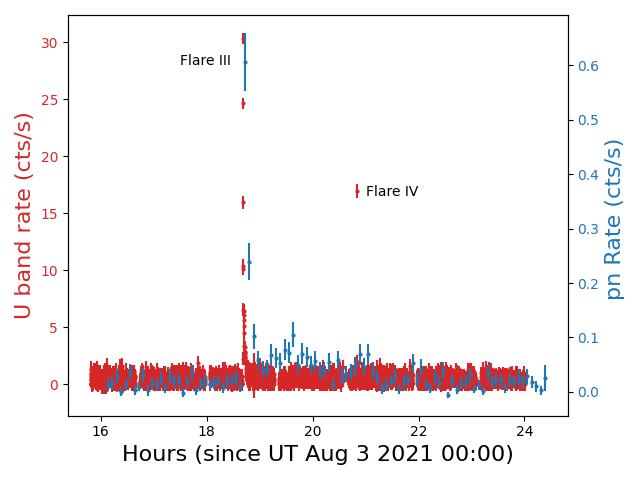}

    \caption{\label{fig:xmm_pn_om}
{\it XMM-Newton} light curve for Teegarden's star obtained with the
OM in the U band (red data points, time resolution 10~s) and 
the EPIC pn detector (blue data points, time resolution 300~s). Flare III occurs at $\sim$18.6\,h and Flare IV occurs at $\sim$20.9\,h.
}
\end{figure}

\subsection{{\it XMM-Newton} observations: Overview}
\label{sec_xmm}

To provide an overview over the whole {\it XMM-Newton} data set for
Teegarden's star, we plot in Fig.~\ref{fig:xmm_pn_om} both
the {\it XMM-Newton} EPIC pn and OM light curves.  Due to the rather different signal strengths different bin sizes were used, namely 
ten~seconds for the U band data and 300~seconds for the
X-ray data.  Both light curves
show a rather low emission level for Teegarden's star.
However, in contrast to {\it Chandra}, the {\it XMM-Newton} X-ray 
data do show a rather major flare (Flare III), followed by some period of
enhanced activity;
the OM on board
{\it XMM-Newton} even shows another  flare (Flare IV), which we  discuss in Sect.~\ref{sec_xmm_flare} in more detail;
the lowest count rates, which we  refer to as quiescence, were encountered at the
beginning and end of the observation.

\subsection{{\it XMM-Newton} observations: Quiescent emission}
\label{sec_xmm_qu}

For Teegarden's star the 4XMM DR13 catalogue lists a total EPIC mean count rate of 
4.16 $\pm$ 0.14 10$^{-2}$ cts/s and 714.6 source counts recorded in the pn detector; 
as obvious from
Fig.~\ref{fig:xmm_pn_om}, the star was in different states during the
{\it XMM-Newton} observations.
For low-flux sources it is  difficult to define a true
"quiescent" level.  Assuming {\it ad hoc} that "quiescence"
applies to the first 6800 seconds and last 8000 seconds
(as in Fig.~\ref{fig:xmm_pn_om}), we find no statistical
differences in the recorded count rates in these two intervals.  We further
estimate that 22\% of the number of total counts were recorded during this 
``quiescent'' state and thus obtain a quiescent count rate of
$2.4 \times 10^{-2}$ cts/s.   Using again PIMMS at NASA's HEASARC \cite{mukai1993}, we
can compute energy flux conversion factors (ECF);
for ''cool" temperatures in the range log~T between 6.2 and 6.5 and
abundances between solar and 0.4 sub-solar, these ECFs change by about
10\%.  Using a flux conversion factor of
$8.57 \times 10^{-13}$ erg\,cm$^{-2}$\,s$^{-1}$ (as appropriate for log~T = 6.2), we derive a
quiescent X-ray flux $f_{\rm X}$ of
$2.06\times 10^{-14}$ erg\,cm$^{-2}$\,s$^{-1}$ and a quiescent X-ray luminosity $L_{\rm X}$ of 
$L_{\rm X}$ = $3.6 \times 10^{25}$ erg\,s$^{-1}$.
We note that these fluxes and luminosities are very close to the
respective values derived from the {\it Chandra} data.

Carrying out a similar exercise for the OM U band data, results in a mean U band rate
of 0.45 $\pm$ 0.02 cts/s.  To convert the observed
U band rates by the {\it XMM-Newton} OM into fluxes,
we use the calibration as presented in ESA's 
{\it XMM-Newton} Calibration documentation
\footnote{\tt{https://www.cosmos.esa.int/web/xmm-newton/calibration-documentation}}.  Specifically,
a conversion factor of 1.98 $\times$ 10$^{-16}$
erg\,cm$^{-2}$\,ct$^{-1}$\,\AA$^{-1}$\, has been derived to convert from observed
rates to flux densities. This is again dominated by systematic errors, which we estimate to be of the order of 10 percent.   With an U band width of
660 \AA \ we can then convert the count rate to U band flux $f_U$ and find
$f_U = 5.9 \times 10^{-14}$ erg\,cm$^{-2}$\,s$^{-1}$ and thus a total U band luminosity $L_{\rm U}$ of
$1.0 \times 10^{26}$ erg\,s$^{-1}$.

The level of X-ray emission in quiescence as measured by {\it Chandra} and \textit{XMM-Newton} of about $L_{\rm X}$ = $4 \times 10^{25}$ erg\,s$^{-1}$ puts 
Teegarden's star among the weakest known stellar X-ray sources.  However, if the relative X-ray luminosity is considered (i.e.  the ratio of $L_X$ to $L_{bol}$), Teegarden's value of $log(L_{\rm X}/L_{bol}) = -4.9$ is much higher than the corresponding solar value, and yet
computing the mean X-ray surface flux $F_X$ by dividing X-ray luminosity $L_X$ and stellar surface area $4\pi R_\star^2$ (cf. Table~\ref{stellpar}), one arrives at values
in the vicinity of the minimal surface flux $F_{X,{\rm min}}$ found by \citet[][Fig.~8]{Schmitt1997A&A...318..215S} for cool main sequence stars;  thus it appears that the minimal flux ``law'' is obeyed down to the lowest mass stars.

\cite{Wright2011ApJ...743...48W} studied the X-ray emission of cool stars as a function of Rossby number (i.e. the ratio of the rotation period to convective turnover time).  Inspecting their Fig.~2 and given the activity levels as measured for Teegarden's star, one expects a Rossby number near unity, implying that the expected rotation period should be equal to the convective turnover time. \cite{Wright2011ApJ...743...48W} also present an empirical determination of the convective turnover time; evaluation of their Eq.~(11) yields a turnover time of $\approx$ 130~days given the mass of Teegarden's star (cf. Table~\ref{stellpar}), which is indeed near the observed rotation period of 97.6\,days \citep{Shan2024}.

\begin{figure} [bth]
\begin{center}
  \includegraphics[width=0.5\textwidth, clip]{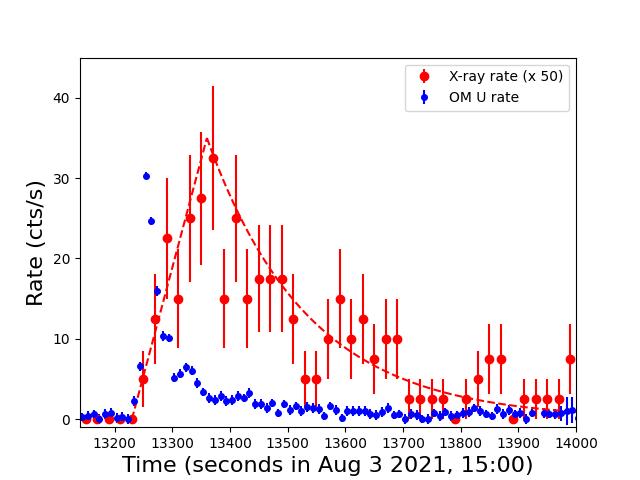}

  \caption{\label{flare_1_higres} Close-up view of \textit{XMM-Newton} data for Flare III: OM U band rate (10 sec bins, blue data points), merged EPIC X-ray rate (20 sec bins, red data points), and flare model curve (dashed line). }
\end{center}
\end{figure}

\subsection{{\it XMM-Newton} observations: Flaring behaviour}
\label{sec_xmm_flare}

The \emph{XMM-Newton} light curve displayed in Fig.~\ref{fig:xmm_pn_om} shows
two episodes of flaring activity, which we examine in more detail in this section.  To this end we employ the highest possible time
resolutions in the data. For the OM this is 1~second, while the
X-ray data are limited by individual photon noise; to obtain the highest S/N, we merge the X-ray data obtained with the three EPIC instruments and consider the recorded photon event arrival times.


In Fig.~\ref{flare_1_higres} we show the {\it XMM-Newton} data recorded for 
Flare III near 13\,250~seconds.  The U band light curve
shows a rapid rise from quiescence to  peak within 10 seconds,
followed by a decay over the next 30~seconds; prior to the main flare rise, there appears to be some
``pre-cursor'' activity lasting for a few seconds.
After the initial rapid flare decay one observes enhanced U band activity, which continues for quite
some time as apparent from Fig.~\ref{fig:xmm_pn_om}.   

An inspection
of the X-ray photon arrival times  shows that during the time of Flare III in the U band  very few X-ray photons  were recorded, while the bulk of the X-ray emission occurs  after the main U band event. 

Specifically, in the seven minutes after the U band Flare III, 127 X-ray photons where recorded, two thirds in the pn-detector and one third
in the two MOS detectors 
with an estimated background contribution of fewer than 10~photons.   Figure~\ref{flare_1_higres} shows that the counting statistics for
the X-ray data is limited. However, from the data we estimate a
lag between the maxima of U band and X-ray emission of about 100~seconds, the same timescale for the X-ray rise, and a decay time of around 175~seconds.  We again note that both the U-band and X-ray emission appear to be enhanced for a while and do not readily return to pre-flare levels.
The lag between U band peak and X-ray peak for Flare III is indicative of the Neupert effect and will be discussed in more detail in Sect.~\ref{sec_neupert}.

\begin{figure} [b]
\begin{center}
  \includegraphics[width=0.5\textwidth, clip]{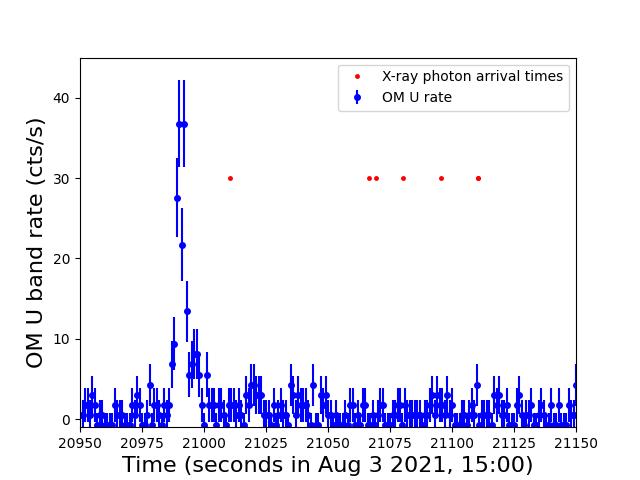}

  \caption{\label{flare_2_higres} Close-up view of \textit{XMM-Newton} data for Flare IV. The OM U band rate (1 sec bins) is shown as blue data points, while the arrival times of the recorded EPIC events are shown as red dots. }
\end{center}
\end{figure}

In Fig.~\ref{flare_2_higres} we show the {\it XMM-Newton} data recorded for  
Flare IV; we note that in contrast to
Fig.~\ref{flare_1_higres}, here the OM U band data are shown with a time resolution of 1~s.   Figure~\ref{flare_2_higres} shows that the
U band flare lasts only about 10 seconds, followed by some enhanced activity
albeit at rather low levels.   Somewhat surprisingly, no significant X-ray signal appears to be
attributable to this UV event.   In Fig.~\ref{flare_2_higres} we plot the arrival times of the recorded X-ray events around the UV event as red points.  During the short U band flare no X-ray events were registered, a first X-ray event was
recorded about 25 seconds after the flare peak, and another ``group'' of five photons about 100~seconds
later.   It is of course possible to interpret this ``group'' as an X-ray response to the U band flare,
on the other hand, judging from the observed X-ray ``background'' rate prior and after the flare, one expects to obtain 2.8 counts in a 125~second
interval; thus the observed 6 photons may equally well be interpreted simply as a statistical fluctuation.  Therefore, to be on the
safe side, we assume the total X-ray output of the flare to be $<$ 6 events.

\subsection{{\it XMM-Newton} observations: Flare energetics}
\label{sec_flare_xmm_ener}

In the following section we focus on the energetics of the observed flaring X-ray and U band emission.  To convert the 
U band rates observed by the \textit{XMM-Newton} OM into fluxes we use the same procedure
as in Sect.~\ref{sec_xmm_qu}, and we list our results in
Table~\ref{flareparxmm}.   Further, to compute bolometric luminosities, we assume a blackbody spectrum for the emitting plasma 
with the same caveats applying; since we do not know the temperature we consider two temperature
values, which hopefully bracket the true values.
\begin{table}
    \caption{\label{flareparxmm} Flare parameters for Teegarden's star from \textit{XMM-Newton} flares}
    \centering
    \begin{tabular}{@{}lrr@{}}
\hline
\hline
Parameter                   & Flare III        & Flare IV\\
\hline
Time of flare peak (U band, UT) & 18:40:50       & 20:49:50\\
Time of flare peak (X-ray, UT)  & 18:42:30       & n.a.\\
Rise time (sec; U band)          & $\approx$ 10             & 5 \\
Rise time (sec; X-ray)           & 100            & n.a. \\
Decay time (sec; U band)         & $\approx$ 20            & 390 \\
Decay time (sec; X-ray)          & 175            & 5 \\
U band rate (at peak, cts/s)           & 55         & 37 \\
U band luminosity (at peak, erg\,s$^{-1}$)    & $1.25 \times 10^{28}$  & $8.4\times 10^{27}$ \\
X-ray rate (EPIC, at peak, cts/s)            & 0.7    & n.a. \\
X-ray luminosity (at peak, erg\,s$^{-1}$)      & $\approx$  $9 \times 10^{26}$         & n.a. \\
Total count U band       & 1488      & 196 \\
Total net count X-ray        & $\approx$ 120         & $<$ 6 \\
Total X-ray fluence (erg)  & $1.6\times 10^{29}$ &  $< 9\times 10^{26}$ \\
Total U band fluence (erg) & $3.4\times 10^{29}$ &  $4.3\times 10^{28}$ \\
\hline
$T_{\rm flare}$ = 15000~K       &                &\\
Peak flare flux (erg\,s$^{-1}$)     & $1.1\times 10^{29}$  &$7.2\times 10^{28}$ \\
Total flare energy (erg)    & $2.9\times 10^{30}$  &$3.8\times 10^{29}$  \\
Flare area (cm$^2$)         & $3.7\times 10^{16}$  &$2.5\times 10^{16}$ \\
\hline
$T_{\rm flare}$ = 8000~K       &                &\\
Peak flare flux (erg\,s$^{-1}$)     & $1.0\times 10^{29}$  &$6.9\times 10^{28}$ \\
Total flare energy (erg)    & $2.8\times 10^{30}$  &$3.7\times 10^{29}$  \\
Flare area (cm$^2$)         & $4.4\times 10^{17}$  & $3.0\times 10^{17}$ \\
\hline
    \end{tabular}
\end{table}

As far as the X-ray energetics are concerned, we note that the derived peak luminosity depends somewhat on the chosen binning; however, 
Fig.~\ref{flare_2_higres} suggests that a value of 0.7 cts/s is
a reasonable value.   Since two thirds of this rate are
due to counts in the pn detector,  using  an ECF of $.11 \times 10^{-12}$ erg\,cm$^{-2}$\,s$^{-1}$ (as appropriate for log~T = 6.8)
leads to a peak X-ray luminosity of  $9 \times 10^{26}$ erg/s and a
flare fluence of 1.6 $\times 10^{29}$ erg. At peak the X-ray output
increased by a factor of more than 20, while the U band output increased by more than a factor of 100.

An inspection of the parameters listed in Table~\ref{flareparxmm} shows that the energy emitted
in the U band alone exceeds the energy emitted at X-ray wavelengths substantially. This discrepancy becomes even larger when the total energy is considered since the U band captures only some part of the emitted bolometric
luminosity. The resulting values for the total emitted energies exceed the corresponding X-ray energies by  more than an
order of magnitude.   While the derived values do by necessity
carry large uncertainties, the conclusion appears inevitable that for
the observed flares the
photospheric energy output is larger by some order of magnitude 
than the coronal output. Such findings are not unusual; for example,
\cite{hawley1995} arrive at similar conclusions for a giant flare observed on
AD~Leo and \cite{kuznetsov2021} find for eight out of nine flares with simultaneous
X-ray and optical coverage that the optical flare output exceeds the corresponding
X-ray output.   The observed ratios between optical and X-ray output can vary substantially,
and it is important to keep in mind that the flare site position on the
stellar surface is normally unknown, and yet the observed optical emission from flares near the
limb can be greatly suppressed (in contrast to X-ray emission).

\begin{figure}[b]
\begin{center}
  \includegraphics[width=0.5\textwidth, clip]{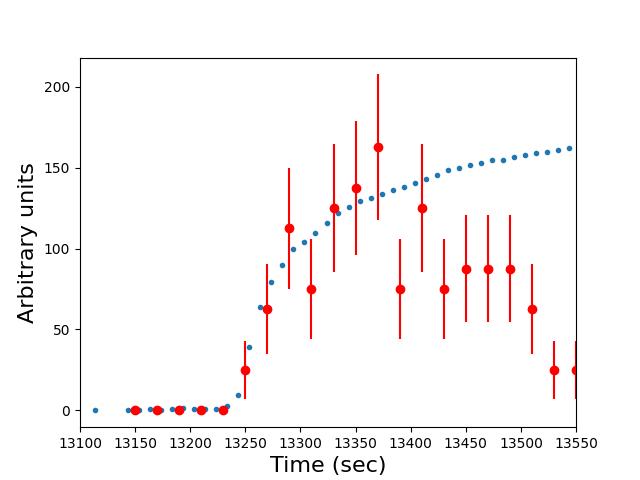}
\caption{\label{neupert} Integrated \emph{XMM-Newton} OM U band data (blue dots) in comparison to X-ray count rate (red data 
points with error bars) for Flare III.}
\end{center}
\end{figure}

\subsection{{\it XMM-Newton} observations: Neupert effect}\label{sec_neupert}
\subsubsection{The Neupert effect in stars}
The  Neupert effect, first described by \cite{neupert1968}, constitutes, according to \cite{kowalski2024} (in chapter 7.7) `the backbone of the solar-stellar flare connection'. The Neupert effect describes an empirically found relationship
between the hard and soft X-ray emission from solar flares.  One often (albeit not always) observes that the observed hard X-ray emission shows the same temporal behaviour as the rate of change of the observed soft X-ray emission.   Such
hard X-ray emission is produced by bremsstrahlung from non-thermal electrons, which travel along the magnetic field lines and finally dissipate their energy in the solar chromosphere and photosphere, producing hot plasma.  The hot plasma expands into the corona and radiates its energy as thermal X-ray radiation.
In many cases, simultaneous hard and soft X-ray observations are not available, and then proxy indicators must be used; for example,
\cite{neupert1968} actually used microwave emission as a proxy for 
non-thermal electrons and their (expected) hard X-ray emission.

In the stellar context, observations of the Neupert effect are relatively rare, and ``hard'' X-ray observations 
(in the sense of solar X-rays) of stellar flares do not exist.  In many cases  simultaneous observations with
the necessary time resolution are not available.
\cite{guedel1996} present simultaneous X-ray observations (using ROSAT and ASCA) and
radio observations (using the VLA at 6~cm and 3.6~cm) of the nearby flare star UV~Cet
and report a few cases of radio bursts followed by soft X-ray flares and argue that
their observations do show a Neupert-like behaviour.
\cite{fuhrmeister2011} report simultaneous X-ray observations (using \textit{XMM-Newton}) and
VLT UVES observations of the nearby flare star Proxima Centauri and demonstrate a
very good agreement between the time derivative of the X-ray light curve and
the optical light curve.
\cite{tristan2023} report multi-wavelength observations of the flare star AU~Mic, using
a variety of different instruments including {\it XMM-Newton} and {\it Swift} in the X-ray range,
the VLA at radio wavelengths and various ground-based observing facilities.  In their data, \cite{tristan2023} find 21 flares with overlapping data coverage,
16 of which are argued to show the Neupert effect.

\subsubsection{Neupert effect in Teegarden's star}
In the case of the {\it XMM-Newton} data of Teegarden's star, we use the
U band emission as proxy indicator for hard X-rays. \cite{qiu2021} show that UV emission is a good proxy indicator
for heating and demonstrate that the observed X-ray emission can be well modelled in its rise; we note, however,
that ``soft'' X-ray emission in a solar context differ from our ``soft'' X-ray data.  In Fig.~\ref{neupert}
we compare the time-integrated U band rate (blue dots) vs. the recorded (and arbitrarily scaled) X-ray flux (red data points with error bars) as a 
function of time for Flare III; we note that the low S/N of our data does not allow us to compute numerical derivatives, we
rather use integrals of the observed UV emission.  As is evident from Fig.~\ref{neupert}, the agreement between these
curves is very good, suggesting that indeed the Neupert
effect is at work.   On the other hand, only six data points
describe the rise of the X-ray emission from the pre-flare level to the peak, thus the agreement might also be
fortuitous to some extent.

\section{Chromospheric activity as observed by CARMENES and ESPRESSO}\label{sec:optical}

\subsection{H$\alpha$ as main chromospheric indicator}\label{sec:halpha}

\subsubsection{General behaviour and flaring activity}
Due to its location in the red wavelength region, where M dwarfs emit most, the H$\alpha$ line at 6564.60 \AA\,(vacuum wavelength, used for all CARMENES wavelengths) has been traditionally the most often used chromospheric line indicator. The $L_{\mathrm{H\alpha}}/L_{\mathrm{bol}}$ 
of Teegarden's star is shown in Fig.~\ref{pEWhalpha} and exhibits strong variability. 

\begin{figure}
\begin{center}
  \includegraphics[width=0.5\textwidth, clip]{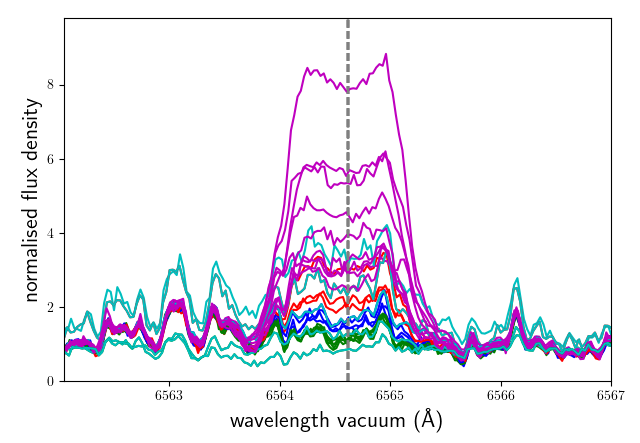}
  \caption{\label{halpha}  Selected spectra of Teegarden's star around the H$\alpha$ line to demonstrate the line shape during quiescent phases and all flares. The spectra of activity level (1) are shown in green, of activity level (2) in blue, and of activity level (3) in red. The flare spectra are shown in magenta. There are a few spectra showing peculiar line shapes with broad additional line components either in absorption or emission (cyan lines). The dashed vertical line marks the central wavelength of H$\alpha$. The normalisation wavelength intervals are located well outside any broad wings at 6537.4--6547.9 and 6577.9--6586.4\,\AA. For the shown spectra the statistical errors are insignificant. 
}
\end{center}
\end{figure}

We show a selection of spectra of Teegarden's star around the
H$\alpha$ line in Fig.~\ref{halpha}. Also directly in the spectra the huge range of variation can be seen. While in the quiescent state the line is more or less absent, during the most active states a strong emission line emerges. In the most quiescent states the line is hard to identify, and also comparison to a PHOENIX purely photospheric spectrum \citep{Husser2013} shows no larger discrepancies at the line position than at other wavelengths. Nevertheless, even the most inactive spectra show some variability at the H$\alpha$ line indicating the presence of chromospheric emission. We show the comparison between observed and model H$\alpha$ line in Fig.~\ref{phoenix}.

When an emission line occurs, it is double horned, though not as pronounced as seen in more active M dwarfs such as G\,080-021 (M3.0\,V, cf. Fig. 3 in \citet{hevar}), which exhibits an H$\alpha$ emission line also during quiescent state. Nevertheless, among the very late-type stars, also vB\,8 exhibits no or only a very shallow self-absorption feature (cf. Fig. 4 in \citet{asym}). Although self-absorption should be symmetric from the theoretical point of view using one dimensional classical chromospheric models \citep{Vernazza1981},
Teegarden's star shows a slightly higher red horn in some spectra. Many stars exhibit one horn preferentially higher than the other; for example, Proxima Centauri also shows a higher red horn for most of the time \citep{proxcen}. This asymmetry in the self-absorption is most probably caused by mass motions, which are not accounted for in classical chromospheric models, but can be included in hydrodynamic simulations of flaring plasma, which then also result in a higher red horn for various evolution stages of the flare \citep{Allred2006}.

For a better description of the variability, we identify from the spectral line shape four different activity states: (1) no line identifiable -- very low activity state, (2) very small line observable -- low activity state, (3) significant emission line -- high activity state, (4) very pronounced emission line -- flaring state. This empirical scheme by eye led to `fuzzy' thresholds in pEW(H$\alpha$)  and we therefore assigned each of these states an (arbitrarily chosen) threshold in pEW and mark them in Fig~\ref{pEWhalpha}. These thresholds are supported by a histogram of the $L_{\rm H\alpha}/L_{\rm bol}$ values shown in Fig.~\ref{histhalpha}.

We believe it quite probable that all spectra of activity state (4) correspond to flares. Then Teegarden’s star as observed with CARMENES in H$\alpha$ is 4.9\% of the time in flaring state, and another 9.7\% in high activity state (3). This flare duty cycle is  only slightly higher than the
3\% found in spectra of the Sloan digital sky survey (SDSS)
for late-type M stars \citep{Hilton2010}. The much lower duty cycle of about 0.007\% in the TESS band is caused by the red continuum wavelengths covered by the TESS band, which needs much higher flare energies involved compared to the H$\alpha$ line to produce a notable flare reaction.  

\subsubsection{Wing asymmetries}\label{sec:asym}

Next to the core line asymmetries, there are several wing asymmetries in the H$\alpha$ line, as seen in Fig.~\ref{halpha} on the blue side of the pink and the brown spectra, which are above the other spectra around 6563 \AA. Such wing asymmetries have been
observed during flares \citep{asym} and we mark these spectra with visually identified wing asymmetries in the pEW time series in Fig.~\ref{pEWhalpha}. 

\begin{figure*}
\begin{center}
  \includegraphics[width=0.49\textwidth, clip]{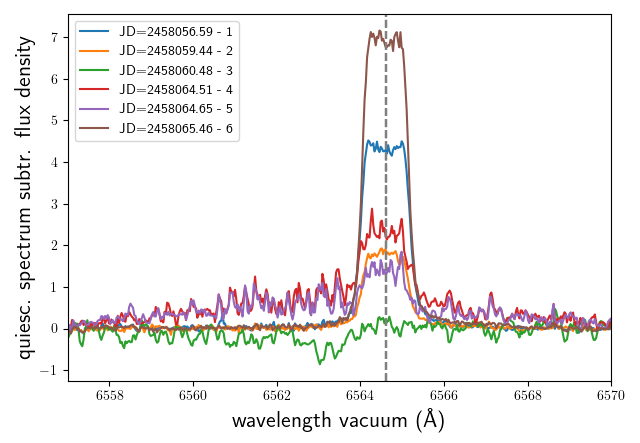}
\includegraphics[width=0.49\textwidth, clip]{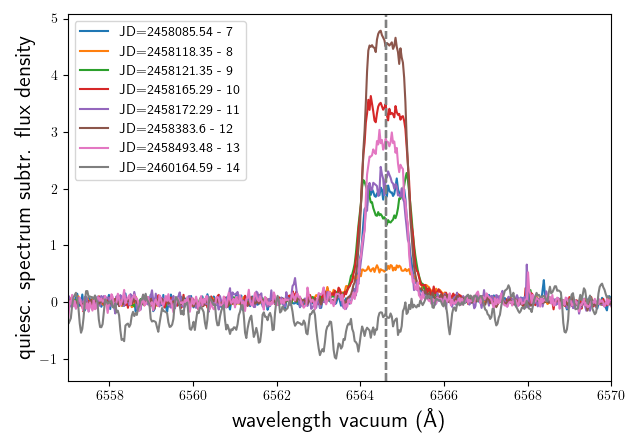}
  \caption{\label{subtracspec}  Flare spectra and spectra with an unusual line shape with a mean
quiescent spectrum subtracted. In the legend the spectra are ordered by   occurrence time, and additionally the label from Fig.~\ref{pEWhalpha} is given. The spectra are split into two panels   for clarity.
}
\end{center}
\end{figure*}

To better demonstrate deviations from the usual line shape, we show all spectra which we have classified as flare spectra or to have an unusual line shape, with the mean quiescent spectrum subtracted, in Fig.~\ref{subtracspec}. 
The flare spectra show a flat top and enhanced amplitudes, but no deviation from the quiescent spectrum outside the main line. The five spectra of unusual line shape, however, show broad wings. One spectrum with low amplitude (JD=2458118.35) only has a symmetric slightly broader shape of unknown origin. Two spectra (JD=2458060.48 and JD=2460164.59) have slight absorption lines in H$\alpha$ and extended absorption wings at the blue side of the line (the latter being the only spectrum with positive  pEW(H$\alpha$)). This may be caused by a prominence rotating at some height with the star. Another two spectra (JD=2458064.51 and JD=2458064.65 in Fig.\ref{subtracspec}) show enhanced wings with more flux extending to lower than to higher wavelength. This suggests that the two spectra are taken during flare onsets, where blue asymmetries are expected due to chromospheric evaporations. The asymmetries are very broad
and we refrain therefore from a Gaussian fit. They extend to about 6570\,\AA\, ($\sim$250 km\,s$^{-1}$) on the red side and to 6558\,\AA\, (\mbox{$\sim -300$ km\,s$^{-1}$}) on the blue side. While the escape velocity of Teegarden's star is about 550 km\,s$^{-1}$,
the two spectra with blue asymmetries are taken by chance in the same night and are only 1.2\,hours apart. Nevertheless, if this is a feature of a flare onset it cannot persist for such a long time (the spectra are taken about 3.4 hours apart). We think it is improbable that two flares happened consecutively each with only its onset covered by the two exposures.  We therefore argue that we may see a coronal mass ejection (CME) under some inclination. 

The two spectra belong to  
a concentration of flaring activity starting
at about JD $=2458056$ followed by another flare three days later
(JD $=2458059$). Again one day later one of the two  spectra showing a broad absorption feature
is taken (JD $=2458060$; denoted by the green line in Fig. \ref{subtracspec}), followed by the two spectra
exhibiting the strong blue emission wing  (JDs $=2458064.51$
and 2458064.65), which precede the strongest
flare covered in our time series at JD $=2458065.46$.
That the possible CME follows one of the possible prominence features may hint at an eruption of the prominence.

\subsubsection{Timing behaviour}\label{sec:timing}

Since we only have snapshot spectra, it is generally not possible to tell, which spectrum of higher activity state is caused by a flare, though decreasing pEW makes this more and more probable, but the threshold chosen here is arbitrary and some of the spectra with a high activity level (3) may be taken during decay phases of large flares, or be small flares themselves. Nevertheless, most of these high-activity pEWs seem to occur during active times longer than the typical duration of a few minutes to hours of a long-lasting H$\alpha$ flare and group to some clusters, which may mark higher activity episodes of Teegarden's star. Surprisingly these clusters are regularly spaced in time and are about 110 to 150 days apart of each other. We mark these clusters in Fig.~\ref{pEWhalpha} and the spacing is the following: 420, 110, 215, 148, 223\,days (the numbers larger than 200\,days are caused by observation gaps, but are about multiples of 110 days).

\begin{figure}
\begin{center}
  \includegraphics[width=0.5\textwidth, clip]{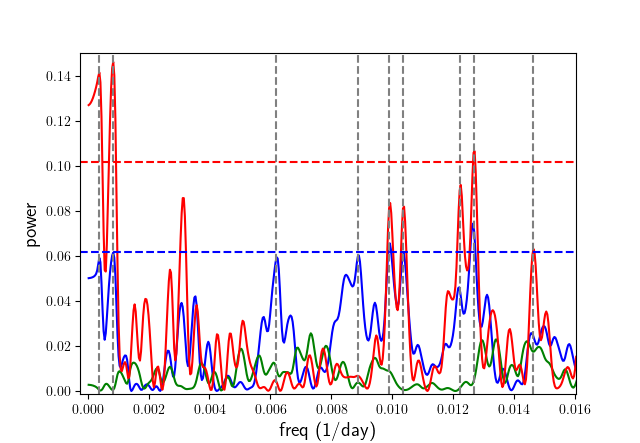}
  \caption{\label{periodogramall} Periodogram of all pEW(H$\alpha$) (blue line) and only activity level (1) and (2) (red line). A periodogram for the shuffled data is shown in green. The dashed blue line corresponds to a FAP level of 10 percent for the power shown in blue, while the dashed red line corresponds to a FAP level of 0.5 percent for the power shown in red. The dashed vertical lines mark the peaks corresponding to periods of 2670, 1200, 161, 112, 100, 96.4, 81.8, 78.8, and 68.4 days (from left to right).}
\end{center}
\end{figure}

To further investigate this issue we have computed a periodogram of all the pEW(H$\alpha$)  and only of the inactive phases (activity level (1) and (2)).
We show the periodograms in Fig.~\ref{periodogramall}. While all the data lead only to insignificant peaks with false alarm probability (FAP) lower than about 10 percent for all peaks, these are predominantly located at the same position as the peaks for the low activity data, which have a much better FAP. Only two peaks seem to stem from the higher activity data since they are absent in the low activity data. These peaks correspond to 112 and 161 days, where 161 days is a one-year alias of 112 days. This strongly supports the regularly cluster spacing of the high-activity data. If this is true, then a stable active region may be the origin of these flares. This would also hint at a higher latitude of this active region if the longer period is caused by solar-like differential rotation (with slower rotation towards the poles) in Teegarden's star.

Further peaks appearing in the periodogram are found at 96 and 100 days, which are aliases of each other regarding the longest period of 2670 days. The peak at 78.8 days corresponds roughly to a one-year alias of 96 days, while the peak at 81.8 days corresponds  to a one-year alias of 112 days. The peak at 1200 days is about half of the longest period peak. Therefore, the 2670-day period remains as the possible activity cycle (see below), as do the 112-day period of the clusters of higher activity and the 96-day period as rotation period  (as is it very close to the 97.56-day period found by \citealt{Shan2024}). Instead,  the 68.4-day period is still unexplained.

Next to the possible periodicity of the occurrence  of the high-activity clusters, there may also be a periodicity present in the low activity pEWs
of H$\alpha$ and \ion{Ca}{ii} IRT of the observing season starting at JD $=2457949$. We show a zoom-in on these data and the periodogram in Figs.~\ref{pEWhalpha2} and \ref{periodogram}. The period we find in these data has a broad peak at a frequency of about 0.0088\,day$^{-1}$ (i.e. at a period of 114\,days). For all other observing seasons the number of observations is too low to do a timing analysis, therefore it may be a quasi-periodic episode, but the clustering of the higher activity states seems to persist on even longer timescales. Both timescales are only slightly higher than the literature rotation period of 97.56\,days \citep{Shan2024}. This seems to indicate, that also during the lowest activity states an activity pattern rotating with the star may be seen, which is veiled for higher activity states (e.g. by dynamically changing H$\alpha$ plages). 

Although this remains speculative, there seems to be a really long variation pattern in the pEW(H$\alpha$) data: an upper envelope of the data seems first to drop and then to rise again (we show a zoom-in on that data in Fig.\ref{pEWhalpha1}), but even without the long data gap more data would be needed here for confirmation and a period determination. Now we can only crudely estimate a period of about 2500\,days ($\sim6.8$\,years), which could be caused by an activity cycle. Already \citet{Dreizler2024} drew a similar conclusion based on mostly the same CARMENES data, but from radial velocity measurements and activity line indices such as differential line width. In the framework of planet detection they found in their Gaussian process analysis a long period of at least the length of the data set and exhibiting a phase shift between radial velocity data and other activity indices, which has been found also for activity cycles on the Sun \citep{CollierCameron2019}.

\subsection{Other chromospheric line indicators}

Next to the H$\alpha$ line we analyse several other atomic lines, which have been used as chromospheric indicators \citep{GdS12, Mittag2018, Patrick, hevar, Fuhrmeister2023, Lafarga2023}. Depending on the spectrograph used, these are  the \ion{Ca}{ii} IRT lines (at 8500.35, 8544.44, and 8664.52\,\AA), the Paschen series including Pa$\beta$, and the \ion{He}{i} IR line ($\sim10830$\,\AA), which are exclusively covered by CARMENES, while the \ion{Ca}{ii} H\& K lines and the Balmer series lines H$\beta$, H$\gamma$, H$\delta$, and H$\epsilon$ are only observed by ESPRESSO. The \ion{He}{i} D$_{3}$ line, the \ion{Na}{i} D doublet, and the \ion{K}{i} doublet at 7700\,\AA, are observed by both spectrographs. 

Generally these lines fall into two types, first, purely chromospheric lines, which are not present in the photosphere, and second, lines, which also have a photospheric absorption line counterpart, which is filled in or shows emission cores. Due to the late M dwarf type, these photospheric absorption lines are quite shallow and very broad for Teegarden's star. Since not many observations of all these chromospheric activity lines for such late and inactive M dwarfs like Teegarden's star exist, we shortly comment on all of them in the following. 

\begin{figure}
\begin{center}
\includegraphics[width=0.5\textwidth, clip]{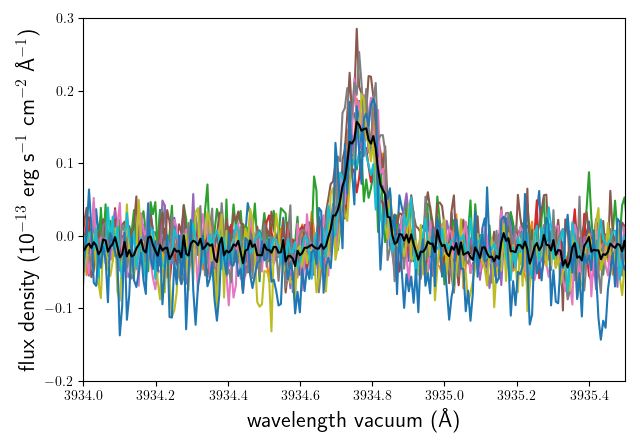}

\includegraphics[width=0.5\textwidth, clip]{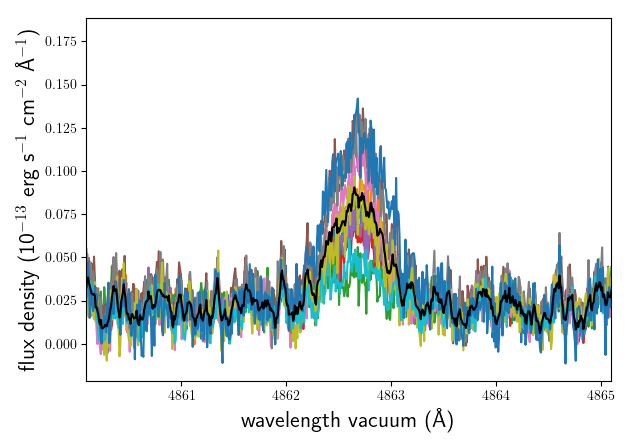}
\caption{\label{cak} All spectra of the \ion{Ca}{ii} K line (\emph{top}) and of the H$\beta$ line (\emph{bottom}) taken with ESPRESSO. The black line denotes the mean spectrum. The \ion{Ca}{ii} K line has no continuum detected.
}
\end{center}
\end{figure}

The \ion{Ca}{ii} H \& K lines are  seen as pronounced emission lines in the ESPRESSO spectra, but the photospheric absorption is not observed, since we do not detect continuum at these very blue wavelengths, as it is increasingly fainter for  M dwarfs of later spectral type. Comparison to a PHOENIX spectrum shows that the continuum is not much lower than our detection limit with the exact numbers depending on the wavelength chosen for normalisation. We show the \ion{Ca}{ii} K line in Fig.~\ref{cak}.

The Balmer series members are all detected except H$\epsilon$, with increasing amplitude to redder wavelength. None of these higher Balmer lines show a double peak structure like H$\alpha$. Since Teegarden's star was in the most quiescent state during this time with little to nearly no H$\alpha$ emission, the detection even of H$\delta$ is a little bit surprising.
We show as an example the H$\beta$ line in Fig.~\ref{cak}.

\begin{figure}
\begin{center}
  \includegraphics[width=0.5\textwidth, clip]{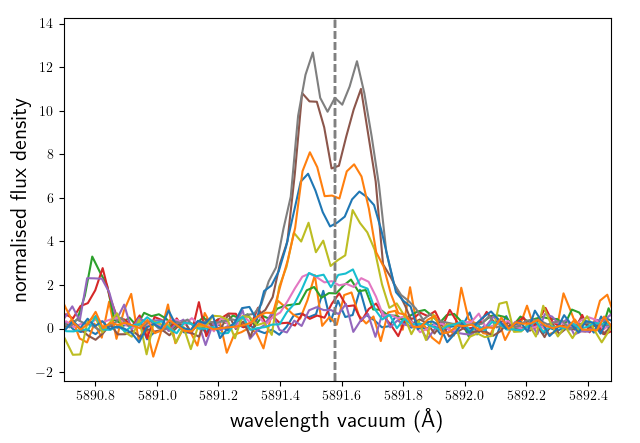}
  \caption{\label{NaD} Selected spectra of Teegarden's star of the blue
   \ion{Na}{i} D line to demonstrate the range of variation in this line.
   We show a few spectra of activity state (1) and all flares except two, which have a low signal-to-noise ratio in the continuum at these wavelength.
}
\end{center}
\end{figure}

The \ion{He}{i} D$_{3}$
line occurs as chromospheric emission line during three of the flares on Teegarden's star and is absent in all other spectra. We show this behaviour in Fig.~\ref{fig:hed3}. The neighbouring \ion{Na}{i} D doublet shows only weak and broad photospheric absorption and also develops emission cores at line centre
during the flares and enhanced activity periods, which we show in Fig.~\ref{NaD}. 

The same is true for the \ion{K}{i} doublet, which exhibits very broad absorption features
contaminated by noise and artefacts from telluric correction. Nevertheless, one can identify some emission cores in the centre of the
photospheric absorption during some of the flares. 

Regarding the \ion{Ca}{ii} IRT lines, the photospheric absorption is not really identifiable for the \ion{Ca}{ii} IRT$_{\rm a}$ line, whose pEW nevertheless shows some variability. For the \ion{Ca}{ii} IRT$_{\rm b}$ line a narrow and weak absorption line can be seen, which fills in for the flares. The \ion{Ca}{ii} IRT$_{\rm c}$ absorption is blended with an \ion{Fe}{i} line at 8664.28\,\AA\, and also develops a fill in for the flaring spectra. We show as an example the middle \ion{Ca}{ii} IRT line in Fig.~\ref{fig:cairt}.

A weak absorption feature is found at the position of the Pa$\beta$ line, which may at least partly be caused by a blend with a presumably molecular feature \citep{Fuhrmeister2023}. No line is found at the position of the \ion{He}{i} IR triplet. Since the \ion{He}{i} D$_{3}$ line is observed as emission line during the largest flares, we would expect some variability for these events also in the \ion{He}{i} IR triplet. The reason that the flares are not seen in the IR line, may be the higher continuum level there than around the \ion{He}{i} D$_{3}$ line.

\begin{figure}
\begin{center}
  \includegraphics[width=0.5\textwidth, clip]{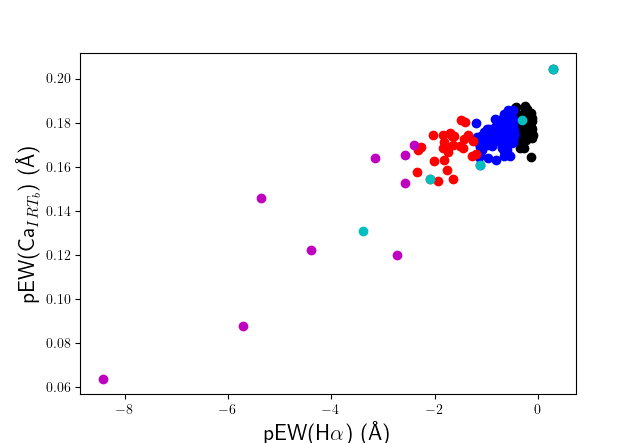}
  \caption{\label{correlation}  Correlation of the pEW of the \ion{Ca}{i} IRT middle line to H$\alpha$. The colours indicate the activity levels as in Fig.~\ref{pEWhalpha}. The typical statistical errors are of the order of one percent and lower for both pEWs.
}
\end{center}
\end{figure}

The pEWs of all of these lines, which are seen also during quiescence, correlate with pEW(H$\alpha$). The highest
correlation is found for the \ion{Ca}{i} IRT$_{\rm b}$ line, with a Pearson's correlation coefficient $r$ of 0.81, while \ion{Ca}{i} IRT$_{a}$ has 0.63,
the blue \ion{Na}{i} D line has 0.47, and the red \ion{K}{i} line has 0.47; all with Pearson's p$<10^{-18}$. We show as an example of these correlations the
one between pEW(H$\alpha$) and pEW(Ca IRT$_{\rm b}$) in Fig.~\ref{correlation}.
The much weaker correlation of the \ion{Na}{i} D and the \ion{K}{i}  line may be at least partly caused by the higher noise level, uncorrected airglow (both especially affect \ion{Na}{i} D) and artefacts from the telluric correction.

\section{Impact of the activity on the habitability of the innermost planets}\label{sec:habitability}

The stellar X-ray and ultraviolet emission is a crucial factor thought to affect planetary habitability \citep[][]{Lammer2009}.
While the planets Teegarden's star b and c may be suitable to harbour liquid surface water under a wide range
of assumptions about the atmosphere \citep[][]{Wandel2019}, it is doubtful whether this is sufficient to support the development
of life. The effects of stellar high-energy irradiation on atmospheric erosion 
\citep[e.g.][]{Sanz2011} and the atmospheric and surface
chemistry on the planets \citep[][]{Ranjan2017, Spinelli2023} must be considered
 as well.   
  
Adopting an X-ray luminosity of $4.2\times 10^{25}$\,erg\,s$^{-1}$ for Teegarden's star (Table~\ref{stellpar}), we
calculate X-ray fluxes at the orbital distances of the planets (Table~\ref{tab:orbitfluxes}). Assuming
a typical value of $10^{27}$\,erg\,s$^{-1}$ for the X-ray luminosity of the Sun \citep[e.g.][]{Peres2000},
we compare these to the
current irradiating X-ray flux of the Earth $f_{\oplus}$. While Teegarden's star does not appear to be overly active
by X-ray standards, the planets b, c, and d all receive higher X-ray fluxes than the Earth. 
In an Earth-like atmosphere, the incoming X-ray flux is absorbed high up in
the atmosphere so that ground-level fluxes are negligible.
The peak X-ray luminosity of the flare observed by {\it XMM-Newton}, exceeds the adopted
quiescent luminosity by about a factor of three (Table~\ref{flareparxmm}), entailing correspondingly
elevated irradiating fluxes. Whether the current irradiation levels and their variability prevent the
evolution of life or are even required to foster chemical processes \citep[][]{Spinelli2023} remains unclear
at this point.

\begin{table}[th]
\caption{X-ray fluxes at planetary orbit distance and comparison to Earth utilising recent X-ray fluxes at Earth $f_{\oplus}$ (above the atmosphere).
\label{tab:orbitfluxes}}
\begin{tabular}{l l l}
\hline \hline
Planet & $f_{\rm X, orbit}$ [erg\,cm$^{-2}$\,s$^{-1}$] & $f_{\rm X, orbit}\,f_{\oplus}^{-1}$ \\
\hline
b & 22.3 & 62.6 \\
c & 7.2 & 20.3 \\
d & 2.4 & 6.7 \\
\hline
\end{tabular}
\end{table}

\section{Comparison to other late M dwarfs}\label{sec:comparison}

Teegarden's star is a rather inactive star for its spectral type, with a quiescent
$\log L_{\mathrm{H\alpha}}/L_{\mathrm{bol}}\sim 5.4$, compared to the typical range of
$\log L_{\mathrm{H\alpha}}/L_{\mathrm{bol}}= -4.8 $ to -4.3 for M6--M7 stars as shown in Fig. 2 by \citet{Liebert2003}. This better compares  to the mean $\log L_{\mathrm{H\alpha}}/L_{\mathrm{bol}}= -4.79$ including also flaring activity of Teegarden's star, and even that would place the star at the lower end of the typical quiescent activity range. This is in line to what one expects for such a slow rotator.

There are only a few comparable flare measurements for late-type M dwarfs. For example, \citet{Fuhrmeister2004} found for the M9 dwarf DENIS 104814.7-395606.1 during a flare a $\log L_{\mathrm{H\alpha}}/L_{\mathrm{bol}}=$ -4.00, which nicely fits the strength of the flares observed for Teegarden's star (see Fig.~\ref{pEWhalpha}).  Also a flare on Proxima~Centauri, which was studied by \citet{proxcen}, led to $\log L_{\mathrm{H\alpha}}/L_{\mathrm{bol}}=$ -4.0. This study also encompassed X-ray observations covering the same flare, which allowed to measure $\log L_{X}/L_{\mathrm{bol}}=$-4.1 and -2.9 during quiescent and flaring state, while for our larger X-ray flare we measure $L_{\rm X}/L_{\mathrm{bol}}=$ -4.38 , compared to a value of $\sim$ -5.0 in quiescence.
This seems to indicate a different activity pattern for both stars, since the values inferred for H$\alpha$ are comparable, while Proxima~Centauri displays much higher X-ray radiation during the quiescent and flaring states. 

For other late-type M dwarfs also more spectacular flares have been found. For example,
\citet{Liebert1999} found for the M9.5
dwarf 2MASSW~J0149090+295613 a quiescent $\log L_{\mathrm{H\alpha}}/L_{\mathrm{bol}}=$-4.63, but during a spectacular flare $\log L_{\mathrm{H\alpha}}/L_{\mathrm{bol}}=$-2.59. Since this flare also exhibited various emission lines in the red part of the spectrum its strength exceeded those of the flares observed on Teegarden's star.  
Another large flare with asymmetries in the \ion{He}{i} infrared triplet line and possibly in the Pa~6 line was studied by \citet{Kanodia2022} for vB~10 (M8). Although the H$\alpha$ line was not covered, its strength could be estimated from the  Ca~IRT and Pa~7, which results in $\log L_{\mathrm{H\alpha}}/L_{\mathrm{bol}}$ estimates ranging from -2.8 to -3.2 for the flare.
While the $\log L_{\mathrm{H\alpha}}/L_{\mathrm{bol}}$ of the quiescent state of these stars is comparable to Teegarden's star, we could not identify such a mega-flare in our data and it stays elusive, if Teegarden's star as relatively inactive star is capable of such strong flares.

Teegarden's star is also interesting as a rare example of a very low-mass and slowly rotating star. It may serve as a testbed for estimating magnetic fluxes from rotation periods as was studied for M dwarfs by \citet{Reiners2022}. For the rotation period of Teegarden's star of 96 days, one arrives at a magnetic flux estimate of 1.7$\times 10^{24}$Mx, using the equation from Table~2 of \citet{Reiners2022}, and one can estimate $L_X$ and $L_{\mathrm{H\alpha}}$ 
from their Fig.~9 leading to $L_X = 6.3\times 10^{26}$\,erg\,s$^{-1}$ and $L_{\mathrm{H\alpha}} = 2.1 \times 10^{26}$\,erg\,s$^{-1}$. 
This is in contrast to our measured $L_{\rm X} = 4.2\times 10^{25}$\,erg\,s$^{-1}$ (see Table~\ref{stellpar}) and $L_{\mathrm{H\alpha}} = 1.1 \times 10^{25}$\,erg\,s$^{-1}$ (from $\log L_{\mathrm{H\alpha}}/L_{\mathrm{bol}}=-5.4$ as our lowest quiescent estimate, see Fig~\ref{pEWhalpha}). Both measured values are about one order of magnitude lower than the expected values. Even, if we use our mean $\log L_{\mathrm{H\alpha}}/L_{\mathrm{bol}}=-4.79\pm0.33$ including flaring activity, there still remains a factor of five discrepancy.
This should indicate, that the estimation of the magnetic flux from rotation period scales differently for these low-mass slow rotators (i.e. it also depends  on mass).

\section{Summary and conclusions}\label{conclusion}

We studied the magnetically induced activity of Teegarden's star by using 298 CARMENES spectra and 11 ESPRESSO spectra taken between 2016 and 2024, covering together all the usually used chromospheric activity indicator lines in the optical range. Moreover, we used three sectors of TESS observations for a search for large flares also affecting the red wavelengths covered by TESS (while smaller flares typically affect only bluer continuum wavelengths). Finally, we analysed one \emph{XMM-Newton} and one \emph{Chandra} observation to assess the X-ray properties of Teegarden's star.

Of the optical chromospheric lines, only the \ion{He}{i} IR triplet is not observed even during flares, which is probably caused by its location at rather red wavelengths, where any chromospheric emission must outshine the strongest part of the photospheric continuum, while the \ion{He}{i} D$_{3}$ line is observed in the spectra of the largest flares. The ESPRESSO spectra allow us to detect \ion{Ca}{ii} H \& K and higher Balmer line emission up to H$\delta$, present also during quiescence. Many other chromospherically influenced lines are present in all observations and exhibit acceptable to good correlation with H$\alpha$, on which we therefore concentrated. 

Our timing analysis of H$\alpha$ exhibited a complex periodic behaviour, showing hints of an activity cycle of the length of the CARMENES observations, which was also found in the RV data \citep{Dreizler2024}. We also identify a peak in the periodogram at about the rotation period in the pEW(H$\alpha$) data and found indications of a periodic repetition of higher activity episodes at a slightly longer period of 112 days. In the lowest activity state of one observing season we also found a periodic behaviour of the same length, hinting at some rotating structure, which is veiled for higher activity states by H$\alpha$ plage variations.

Thus, all data taken together paint the picture of a generally  inactive star, which is nevertheless capable of producing substantial flaring. In pEW(H$\alpha$) we observe the star to be 4.9\% of the time in a flaring state, which is slightly higher than what \citet{Hilton2010} observed for late M dwarfs in the SDSS. Moreover, in the H$\alpha$ line shape and its asymmetries we find indications for a prominence erupting and maybe even causing a CME. In the X-ray range we also find large flares and we observe the Neupert effect in one flare, suggesting the production of non-thermal hard X-ray emission. Finally, the flare energetics derived from the
X-ray and TESS data place these flares among large solar flares.

\begin{acknowledgements}
We acknowledge the help of J-U.~Ness with the \textit{XMM-Newton} OM data.
We thank J. Sanz-Forcada for some discussions.
  This publication was based on observations collected under the CARMENES Legacy+ project.
  CARMENES is an instrument at the Centro Astron\'omico Hispano en Andaluc\'ia (CAHA) at Calar Alto (Almer\'{\i}a, Spain), operated jointly by the Junta de Andaluc\'ia and the Instituto de Astrof\'isica de Andaluc\'ia (CSIC).
  The authors wish to express their sincere thanks to all members of the Calar Alto staff for their expert
  support of the instrument and telescope operation. 
  CARMENES was funded by the Max-Planck-Gesellschaft (MPG), 
  the Consejo Superior de Investigaciones Cient\'{\i}ficas (CSIC),
  the Ministerio de Econom\'ia y Competitividad (MINECO) and the European Regional Development Fund (ERDF) through projects FICTS-2011-02, ICTS-2017-07-CAHA-4, and CAHA16-CE-3978, 
  and the members of the CARMENES Consortium 
  (Max-Planck-Institut f\"ur Astronomie,
  Instituto de Astrof\'{\i}sica de Andaluc\'{\i}a,
  Landessternwarte K\"onigstuhl,
  Institut de Ci\`encies de l'Espai,
  Institut f\"ur Astrophysik G\"ottingen,
  Universidad Complutense de Madrid,
  Th\"uringer Landessternwarte Tautenburg,
  Instituto de Astrof\'{\i}sica de Canarias,
  Hamburger Sternwarte,
  Centro de Astrobiolog\'{\i}a and
  Centro Astron\'omico Hispano-Alem\'an), 
  with additional contributions by the MINECO, 
  the Deutsche Forschungsgemeinschaft through the Major Research Instrumentation Programme and Research Unit FOR2544 ``Blue Planets around Red Stars'', 
  the Klaus Tschira Stiftung, 
  the states of Baden-W\"urttemberg and Niedersachsen, 
  and by the Junta de Andaluc\'{\i}a.
  We acknowledge financial support from the Agencia Estatal de Investigaci\'on (AEI/10.13039/501100011033) of the Ministerio de Ciencia e Innovaci\'on and the ERDF ``A way of making Europe'' through projects 
  PID2021-125627OB-C31,         
  PID2019-109522GB-C5[1:4],     
  and the Centre of Excellence ``Severo Ochoa'' and ``Mar\'ia de Maeztu'' awards to the Instituto de Astrof\'isica de Canarias (CEX2019-000920-S), Instituto de Astrof\'isica de Andaluc\'ia (SEV-2017-0709) and Institut de Ci\`encies de l'Espai (CEX2020-001058-M).
  This work was also funded by the Generalitat de Catalunya/CERCA programme and the Agencia Estatal de Investigaci\'on del Ministerio de Ciencia e Innovaci\'on (AEI-MCINN) under grant PID2019-109522GB-C53 and by the Deutsche Forschungsgemeinschaft under grant DFG SCHN~1382/4-1
  This work made use of PyAstronomy \citep{pya}, which can be downloaded at {\tt https://github.com/sczesla/PyAstronomy}.

\end{acknowledgements}

\bibliographystyle{aa}
\bibliography{ref}

\appendix
\section{Chromospheric activity}
We show in this section additional CARMENES data demonstrating the different chromospheric activity levels and involved timescales of Teegarden's star.

\begin{figure}
\begin{center}
  \includegraphics[width=0.5\textwidth, clip]{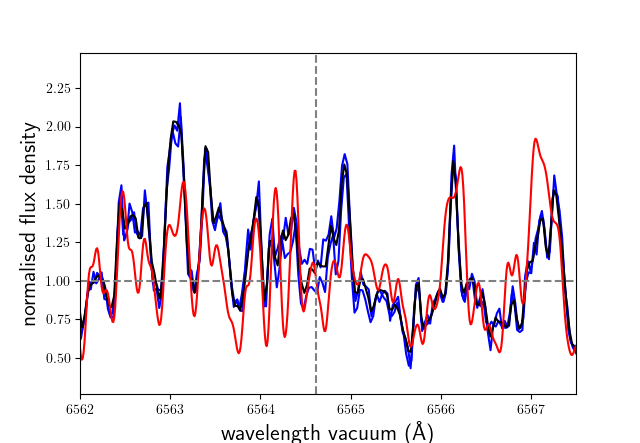}
  \caption{\label{phoenix} Comparison of the mean quiescent observed spectrum of Teegarden’s star (black line) to a PHOENIX photospheric model spectrum
        (red line) with $T_{\rm eff} =3000$\,K and $\log g= 5.0$ around H$\alpha$. Additionally we show two arbitrarily chosen spectra of low activity state (blue lines) demonstrating variability at the line position. While the dashed vertical line marks the position of the H$\alpha$ line, the dashed horizontal line marks the normalisation level.
}
\end{center}
\end{figure}

Regarding the lowest activity state of Teegarden's star, we compare a median spectrum of all spectra in activity level (1) with a model photospheric spectrum from the PHOENIX \citep{Hauschildt1999}
model spectrum library \citep{Husser2013}. We overplot here a model spectrum with
$T_{\rm eff} =3000$\,K and $\log g = 5.0$ and solar chemical composition and show the comparison between model and observed spectrum in Fig.~\ref{phoenix}. Although there are differences in the details, the overall spectral appearence is resembled quite well by the model, given that molecular data is notoriously badly known. Also at the location of the H$\alpha$ line there are no larger deviations from the photospheric model seen.

\begin{figure}
\begin{center}
  \includegraphics[width=0.5\textwidth, clip]{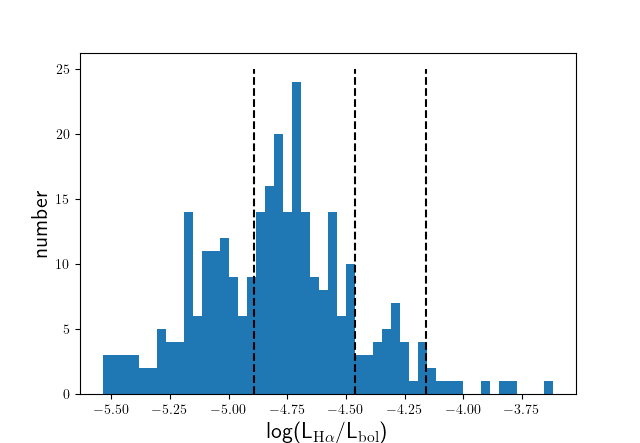}
  \caption{\label{histhalpha} Histogram of the $L_{\rm H\alpha}/L_{\rm bol}$ values. The thresholds between the (empirically defined) different activity states of Teegarden's star are shown as dashed vertical lines.
}
\end{center}
\end{figure}

In Fig.~\ref{histhalpha} we show a histogram of the $L_{\rm H\alpha}/L_{\rm bol}$  values of Teegarden's star. At about $L_{\rm H\alpha}/L_{\rm bol}$=-4.9 there are lower numbers of spectra detected (which can be seen as gap most clearly in the observing season of 2016 in Fig.~\ref{pEWhalpha}), which co-incises with our first threshold between activity level (1) and (2).   

\begin{figure}
\begin{center}
  \includegraphics[width=0.5\textwidth, clip]{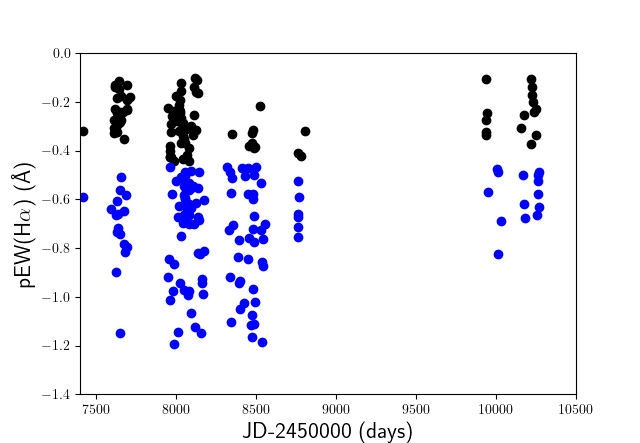}
  \caption{\label{pEWhalpha1} Time series of pEW(H$\alpha$) as in Fig.~\ref{pEWhalpha}. We mark low activity states as black and blue dots, higher activity states as  red dots, and spectra with an uncommon H$\alpha$ shape with cyan dots. The upper envelope of the black dots show first an increase, then a decrease and again high pEW(H$\alpha$) values for the last data points.
}
\end{center}
\end{figure}

\begin{figure}
\begin{center}
  \includegraphics[width=0.5\textwidth, clip]{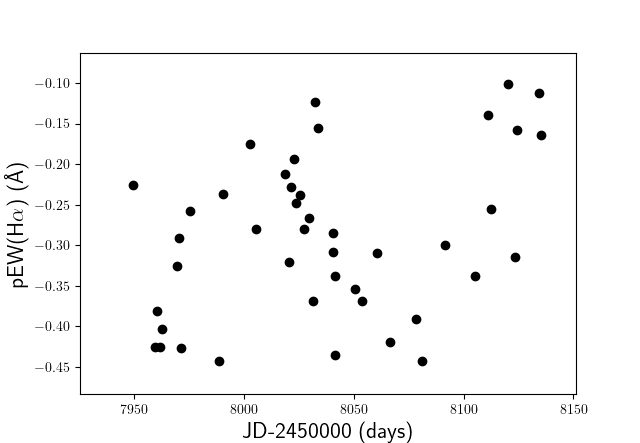}
  \caption{\label{pEWhalpha2} Time series of the most inactive pEW(H$\alpha$) of one observing season. 
}
\end{center}
\end{figure}

\begin{figure}
\begin{center}
  \includegraphics[width=0.5\textwidth, clip]{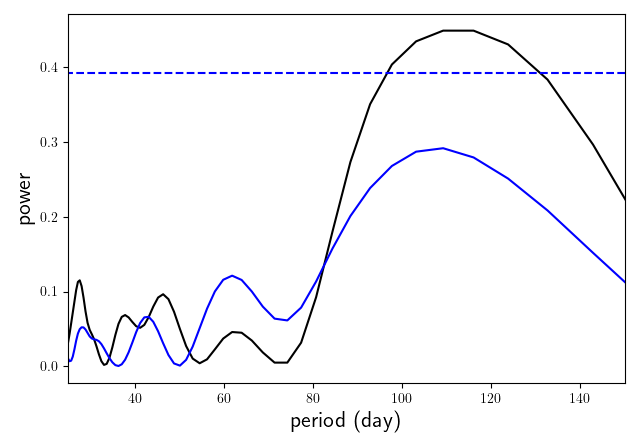}
  \caption{\label{periodogram} Periodogram of the pEW(H$\alpha$) (black line) and pEW(Ca IRT 8544\AA) (blue line). The dashed horizontal line marks the FAP = 0.01 level.}
\end{center}
\end{figure}

To make some of the features in Fig.~\ref{pEWhalpha}  clearer, we show in Fig.~\ref{pEWhalpha1} a zoomed-in image of the low activity pEW(H$\alpha$)s to make the long-term variability of the upper envelope (but also of the lower envelope of the higher activity data marked in blue dots; however, this is not as clearly seen) visible more easily. In Fig.~\ref{pEWhalpha2} we zoom-in even further on the lowest activity pEW(H$\alpha$) data in one observing season, to show the periodic behaviour seen there. In Fig.~\ref{periodogram} we show the corresponding periodogram.

\begin{figure}
\begin{center}
  \includegraphics[width=0.5\textwidth, clip]{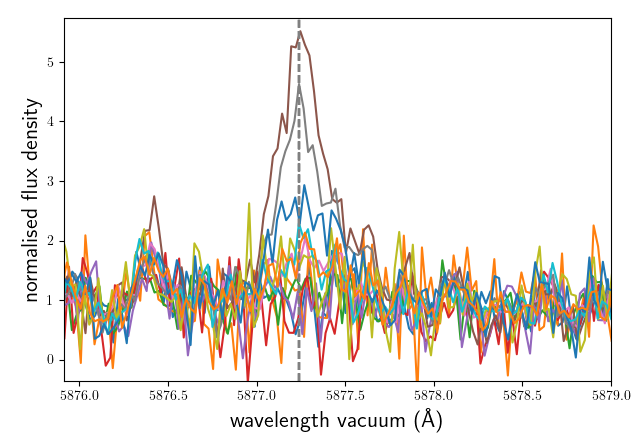}
  \caption{\label{fig:hed3}   Selected CARMENES spectra of Teegarden's star of the \ion{He}{i} D$_{3}$ line. The shown spectra are the same as in Fig.~\ref{NaD}.
}
\end{center}
\end{figure}

\begin{figure}
\begin{center}
  \includegraphics[width=0.5\textwidth, clip]{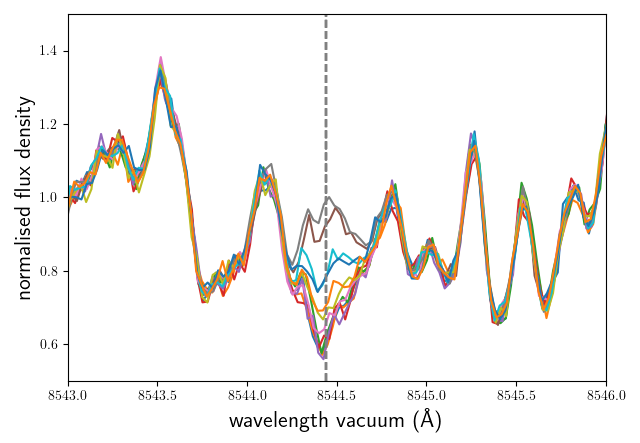}
  \caption{\label{fig:cairt}   Selected CARMENES spectra of Teegarden's star of the  middle \ion{Ca}{ii} IRT line. The
line shows a fill in during flares. The shown spectra are the same as in Fig.~\ref{NaD}.
}
\end{center}
\end{figure}

As further examples of chromospherically active lines we show in Fig.~\ref{fig:hed3} the \ion{He}{i} D$_{3}$ line and in Fig.~\ref{fig:cairt}
the middle line of the \ion{Ca}{ii} IRT.

\end{document}